\documentclass{article}
\usepackage[margin=0.8in]{geometry}

\RequirePackage[OT1]{fontenc}
\RequirePackage{amsthm,amsmath,amsfonts,amssymb}
\RequirePackage{natbib}
\usepackage{float, graphicx, caption}
\usepackage{pgf,tikz}
\usetikzlibrary{positioning}
\usetikzlibrary{arrows.meta, matrix}

\usepackage{tablefootnote}
\usepackage{enumerate, enumitem}
\usepackage{url} 

\usepackage[nokeyprefix]{refstyle}
\usepackage{varioref}
\usepackage{xr-hyper}
\RequirePackage[colorlinks,citecolor=blue]{hyperref}

\usepackage{titlesec}
\titleformat{\subsubsection}[runin]
        {\normalfont\bfseries}
        {\thesubsubsection}
        {0.5em}
        {}
        [.]

\titleformat{\section}
        {\large\bfseries}
        {\thesection}
        {0.5em}
        {}
        []

\titleformat{\subsection}
        {\normalfont\bfseries}
        {\thesubsection}
        {0.5em}
        {}
        []

\usepackage{bbold}

\usepackage{bm}

\usepackage{subcaption}

\usepackage{setspace}

\usepackage{longtable}
\usepackage{array}
\newcolumntype{L}[1]{>{\raggedright\let\newline\\\arraybackslash\hspace{0pt}}m{#1}}
\newcolumntype{C}[1]{>{\centering\let\newline\\\arraybackslash\hspace{0pt}}m{#1}}
\newcolumntype{R}[1]{>{\raggedleft\let\newline\\\arraybackslash\hspace{0pt}}m{#1}}

\newcommand*{\mybullet}{\raisebox{-0.25ex}{\scalebox{1.8}{$\cdot$}}}

\newcommand{\tend}[1]{\hbox{\oalign{$\bm{#1}$\crcr\hidewidth$\scriptscriptstyle\bm{\sim}$\hidewidth}}}
\newcommand\numberthis{\addtocounter{equation}{1}\tag{\theequation}}

\newcommand{\mb}{\bm}
\newcommand{\independent}{\perp\!\!\!\perp}
\newcommand{\notindependent}{\not\!\perp\!\!\!\perp}

\newcommand{\PM}{PM$_{2.5}$}

\usepackage[capitalise,noabbrev]{cleveref}

\allowdisplaybreaks

\begin{document}
	
\crefformat{equation}{#2\textup{(#1)}#3}
\crefformat{section}{\S#2#1#3}
\crefformat{subsection}{\S#2#1#3}
\crefformat{subsubsection}{\S#2#1#3}

\title{A causal exposure response function with local adjustment for confounding: Estimating health effects of exposure to low levels of ambient fine particulate matter}
\author{Georgia Papadogeorgou$^1$\thanks{Funding for this work was provided by National Institutes of Health R01 ES024332, USEPA 83587201-0, and Health Effects Institute 4953-RFA14-3/16-4.} \hspace{2pt}
and Francesca Dominici$^2$\\[5pt]
{\small $^1$Department of Statistical Science, Duke University, Durham NC 27710} \\
{\small $^2$Department of Biostatistics, Harvard T.H. Chan School of Public Health, Boston MA 02115 }}
\date{}
\maketitle

\begin{abstract}

In the last two decades, ambient levels of air pollution have declined substantially. At the same time, the Clean Air Act mandates that the National Ambient Air Quality Standards (NAAQS) must be routinely assessed to protect populations based on the latest science.
Therefore, researchers should continue to address the following question: is exposure to levels of air pollution below the NAAQS harmful to human health? Furthermore, the contentious nature surrounding environmental regulations urges us to cast this question within a causal inference framework.
Several parametric and semi-parametric regression approaches have been used to estimate the exposure-response (ER) curve between long-term exposure to ambient air pollution concentrations and health outcomes. 
However, most of the existing approaches are not formulated within a formal framework for causal inference, adjust for the same set of potential confounders across all levels of exposure, and do not account for model uncertainty regarding covariate selection and the shape of the ER.

In this paper, we introduce a  Bayesian framework for the estimation of a causal ER curve called LERCA (Local Exposure Response Confounding Adjustment), which a) allows for different confounders \textit{and} different strength of confounding at the different exposure levels; and b) propagates model uncertainty regarding confounders' selection and the shape of the ER. Importantly, LERCA provides a principled way of assessing the observed covariates' confounding importance at different exposure levels, providing researchers with important information regarding the set of variables to measure and adjust for in regression models.
Using simulation studies, we show that state of the art approaches perform poorly in estimating the ER curve in the presence of local confounding.

LERCA is used to evaluate the relationship between long-term exposure to ambient \PM{}, a key regulated pollutant, and cardiovascular hospitalizations for 5,362 zip codes in the continental U.S. and located near a pollution monitoring site, while adjusting for a potentially varying set of confounders across the exposure range. Our data set includes rich health, weather, demographic, and pollution information for the years of 2011-2013.
The estimated exposure-response curve is increasing indicating that higher ambient concentrations lead to higher cardiovascular hospitalization rates, and ambient \PM{} was estimated to lead to an increase in cardiovascular hospitalization rates when focusing at the low exposure range. Our results indicate that there is no threshold for the effect of \PM{} on cardiovascular hospitalizations.

\end{abstract}

\textit{keywords:} air pollution, cardiovascular hospitalizations, causal inference, exposure response function, local confounding, low exposure levels, particulate matter


\section{Introduction}
The Clean Air Act, one of the most comprehensive and expensive air quality regulations in the world, mandates that the National Ambient Air Quality Standards (NAAQS) are routinely reviewed. If evidence of the adverse health effects of exposure to ambient air pollution at levels below the NAAQS is established based on the peer reviewed literature, then the NAAQS must be lowered, even at the cost of hundreds of million of dollars. For that reason, researchers routinely address the following question: is exposure to levels of air pollution, even below the NAAQS, harmful to human health?
With the next review of the NAAQS for fine particulate matter (PM$_{2.5}$) scheduled to be completed by the end of the year 2020, the determination of whether exposure levels of PM$_{2.5}$ below the NAAQS is harmful to human health is subject to unprecedented level of scrutiny. More recently, because of the highly contentious nature surrounding air pollution regulations and the lowering of the NAAQS particularly, there is an increasing pressure to cast this question within a causal inference framework \citep{Zigler2014a}. The method in this paper is motivated by the need to address this critically important question by flexibly estimating an exposure response curve while reliably eliminating confounding bias especially \textit{at low levels} of exposure. 

The literature on the harmful effects of air pollution is very extensive (see, for example, \cite{Dominici2002, Eftim2008, Zeger2008, Zanobetti2007,Crouse2015, Crouse2016, Di2017association, Di2017air, Berger2017, Makar2018, Lim2018association}). However, significant methodological gaps remain in the context of estimating health effects at very low levels.
Environmental research studying the health effects of exposure to low levels of ambient air pollution has either examined the relationship in the subset of the sample residing in areas with ambient concentrations below a pre-specified threshold \citep{Lee2016acute, Shi2016, Di2017association, Di2017air, Schwartz2017estimating, Makar2018, Wang2018longterm, Schwartz2018national}, or has employed regression approaches for ER estimation across the observed range of pollution concentrations \citep{Daniels2000, Dominici2002, Schwartz2002, Bell2006, Hart2015association, Thurston2016ambient, Jerrett2017comparing, Weichenthal2017biomass, Lim2018association}.
In either case, confounding adjustment in air pollution studies is most-often performed using either a pre-specified set of covariates, or a set of covariates which is decided upon using an ad-hoc variable selection procedure. Such procedure is often based on the statistical significance of covariates' coefficients in an outcome regression, or the change in the pollution concentration's coefficient in an outcome model including and excluding sets of covariates \citep{Devries2016, Pinault2016, Garcia2016, Weichenthal2017biomass}.

Generally, regression and semi-parametric modeling approaches for ER estimation such as generalized linear models or generalized additive models \citep{Hastie1986, Daniels2004, Shaddick2008, Shi2016, Dominici2002} make the following assumptions:
1) the set of potential confounders that are included into the regression model among a potentially large set of available covariates is specified a priori;
2) uncertainty arising from the variable selection techniques is not accounted for;
3) the same potential confounders with constant confounding strength are considered when estimating the health effects across all exposure levels (we refer to this as \textit{global confounding adjustment}); 
and 4) the shape of the ER function is modelled as a spline, a polynomial, or linear with a threshold.

Even though ER estimation in air pollution research has mostly remained outside the potential outcome framework, there has been substantial work in ER estimation within the causal inference literature. \cite{Hirano2004} introduced the generalized propensity score (GPS) in order to adjust for confounding when estimating the effects of a continuous exposure. \cite{Flores2012estimating} estimated a causal ER function employing a weighted locally linear regression with weights defined based on the GPS.  \cite{Kennedy2017} introduced a doubly robust approach for estimating the causal ER function using flexible machine learning tools. 

These approaches are very promising and manifest the growing interest in principled causal inference methods for continuous exposures. However, none of the existing approaches explicitly accommodates that in ER estimation, and in contrast to binary treatments, confounding \textit{might differ across levels of the exposure}. In fact, even though some of the approaches could be altered to allow for different set of confounders or different confounding strength across exposure levels, current implementations of causal methodology for ER estimation has assumed global confounding of pre-selected covariates. Furthermore, it is unclear how these approaches perform in the case of confounding that varies across exposure levels. To address this, confounding adjustment and confounder selection need to be meaningfully extended in the case of a continuous exposure to provide useful scientific guidance with regard to covariates' confounding importance \textit{at different exposure levels}.

In our exploratory analyses (\cref{sec:data_description}), we report that the relationship between ambient PM$_{2.5}$ concentrations and the rate of hospitalization for cardiovascular diseases might be confounded by a {\it different set of covariates} at the low versus at the high exposure levels, or by covariates with \textit{different confounding strength}. We refer to this phenomenon as \textit{local confounding}. We argue that --especially in the context of estimating causal effects at low levels-- local confounding adjustment is deemed necessary.

To target local confounding, if exposure levels with different confounding were known, one could adopt a separate model at each level and adjust for \textit{all} measured variables using one of the approaches described above. However, even if the number of covariates and local sample size rendered such approach computationally feasible, including unnecessary confounders in the regression model could lead to inefficient estimation of causal effects, especially at very low levels of exposure where data are sparse. Data driven methods to select a minimal necessary set of covariates to be included into an outcome model for estimation of causal effects of binary treatments have been proposed \citep{Luna2011a, Wang2012, Wilson2014}, but to our knowledge, they have not been extended in the context of ER estimation with local confounding adjustment.


The goal of this paper is to overcome the challenges described above by introducing 
a Bayesian framework for the estimation of a causal ER curve called LERCA (Local Exposure Response Confounding Adjustment).
We cast our approach within a causal inference framework by introducing the concept of {\it experiment configuration} 
$\mb{\bar{s}} = (s_0, s_1, \dots, s_{K + 1})$, where $[s_{k-1}, s_k)$ denotes a specific range of exposure values.
We use the term {\it experiment} to mimic the hypothetical assignment of a unit to exposure value within $[s_{k-1}, s_k)$. Within each experiment, i.e. \textit{locally} in the exposure range $[s_{k-1}, s_k)$, we assume that: 1) the ER is linear; 2) the potential confounders of the ER relationship are unknown but observed; and 3) the strength of the local confounding is also unknown. Across experiments, we require that the ER is continuous at the points $\mb{\bar s}$. Importantly, the internal points of the experiment configuration, $s_1, s_2, \dots, s_K$, are themselves unknown and have to be estimated from the data.

Our work contributes to various components in the literature. First, we contribute to the estimation of causal effects of continuous treatments by extending our understanding of confounding in these settings. Second, our work has connections to the literature on Bayesian free-knot splines \citep{Denison1998automatic, Dimatteo2001bayesian}. The location of the knots (internal points of the experiment configuration) is informed by both the ER fit, and the necessity for local confounding adjustment. Lastly, our work contributes to the highly controversial and politically charged issue of estimating the causal effects of population exposure to low levels of ambient air pollution.

Even though our motivation and focus is the effects of air pollution, the statistical challenges related to ER estimation at low exposure levels are common across many fields, such as toxicology \citep{Scholze2001}, and clinical trials \citep{Babb1998}. 
In fact, the methodology presented in this paper can be used to evaluate regulatory settings of potential harmful substances, and can be routinely used to assess health effects of low level exposures.
Such applications include the effects of lead \citep{Chiodo2004, Jusko2008}, environmental contaminants \citep{vanderoost2003}, radiation \citep{National2006, Fazel2009}, and pesticides \citep{Mackenzie2010, Androutsopoulos2012}.

In \cref{sec:data_description} we introduce our motivating data set, discuss the difference between personal exposures and ambient concentrations in air pollution studies, and illustrate that local confounding is likely to be present in our study. In \cref{sec:notation}, we introduce the notation and assumptions on which LERCA in \cref{sec:method} is based. In \cref{sec:simulations} we show through simulations that both off-the-shelf and state of the art approaches for ER estimation perform poorly when local confounding is present, and we compare LERCA to alternatives in the presence of global confounding. Finally, in \cref{sec:application}, we use LERCA to estimate the causal ER function relating ambient PM$_{2.5}$ concentrations with log cardiovascular hospitalization rates in the Medicare population of 5,362 zip codes. Limitations and potential extensions are discussed in \cref{sec:discussion}.

\section{Data description, ambient concentrations, local confounding}
\label{sec:data_description}

In this section we illustrate that, in our study, there might exist a different set of confounders at the low and the high levels of ambient pollution concentrations. LERCA is motivated to overcome this particular challenge.

\subsection{Data description}

\begin{figure}[!t]
	\begin{center}
	\includegraphics[width = 0.65\textwidth]{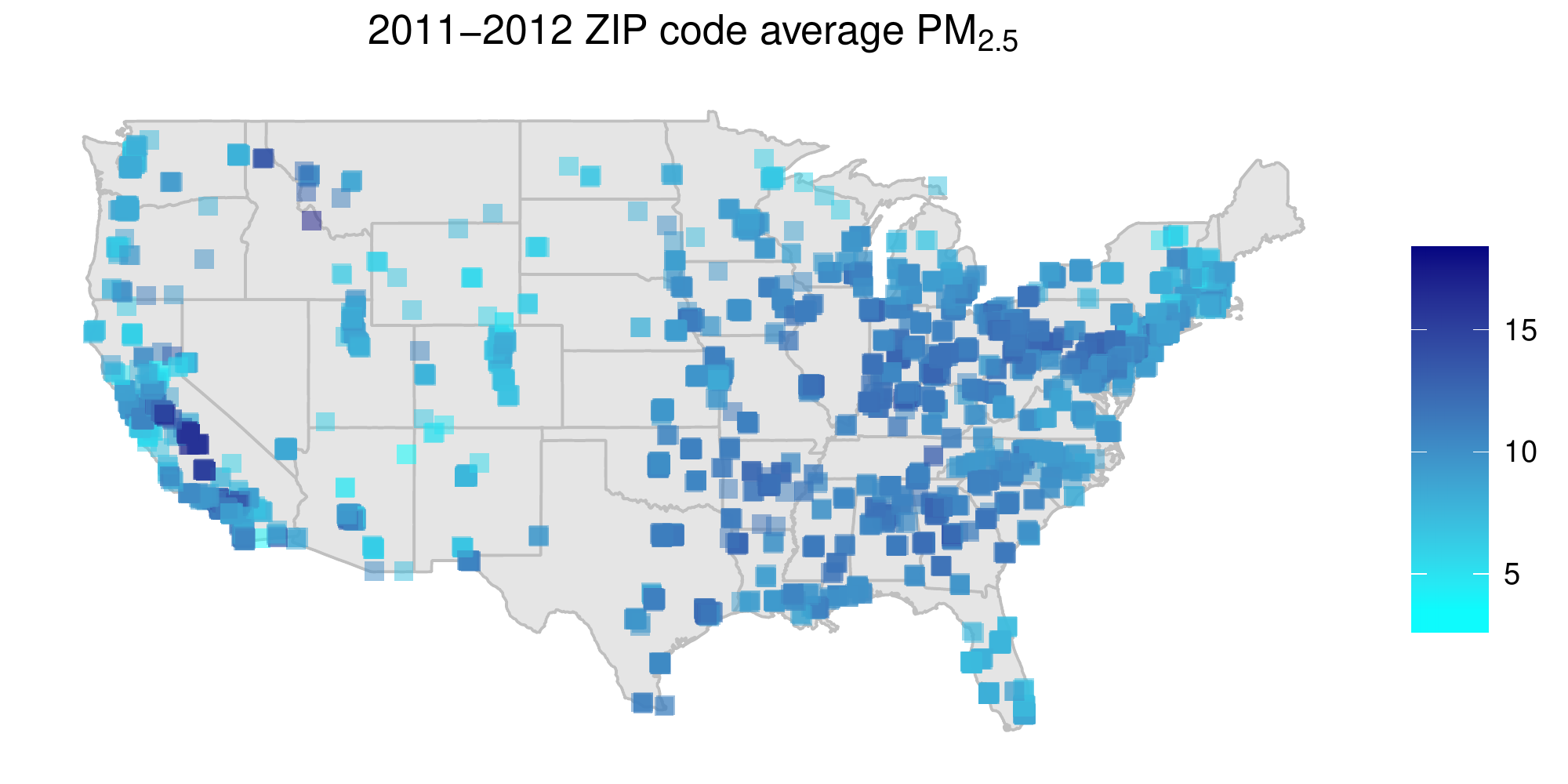}
	\caption{Average levels of PM$_{2.5}$ for the years 2011-2012 for each zip code $i$ included into the analysis.}
	\label{fig:PM_2013}
	\end{center}
\end{figure}

We start by briefly describing our data set which is a collection of linked data from many sources. The unit of the observation is the zip code $i$, with sample size $N= 5,362$. For each zip code,
we acquired information on several potential confounders, denoted by $C_{ij}$ for  $j=1, 2, \ldots, p$ and $p=27$, capturing socio-economic, demographic, climate, and risk factor information. The full set of zip code level covariates are described in Table \ref{app_tab:Table1}.
We calculate the outcome $Y_i$ defined as log hospitalization rate for cardiovascular diseases (codes ICD-9 390 to 459) among Medicare beneficiaries residing in zip code $i$ in the year 2013.
Since Medicare beneficiaries are, in their plurality, individuals over the age of 65, our focus is on the health effect of \PM{} on the elderly. For each zip code $i$, we assign exposure $X_i$ defined as the average of daily levels of ambient PM$_{2.5}$ concentrations for the years 2011 and 2012 recorded by EPA (U.S. Environmental Protection Agency) monitors within a 6 mile radius of zip code $i$'s centroid.
The values of $X_i$ range from $2.7$ to $18.3 \mu g \slash m^3$ (see \cref{fig:PM_2013}).
We define $X_i$ using the two years prior to the year whose outcome we analyze in order to respect the temporal ordering of treatment and outcome when drawing causal conclusions. Longer time lags could be considered, but, in such settings, our analysis would potentially be more susceptible to population mobility.

Since our definition of $X_i$ requires the presence of an EPA monitor within 6 miles of a zip code's centroid, the zip codes included in our study are a subset of the full set of zip codes in the continental U.S. Supplement \ref{sec:data_details} includes a detailed description of data linkage (EPA monitors, Medicare, others), and descriptive statistics across zip codes in the whole continental U.S. and only those included in our study. Excluded zip codes resemble, in general, those included in our analysis, but are perhaps in more rural areas, with lower population density, and higher proportions of white population and unemployment.

\subsection{Ambient concentrations versus personal exposures}

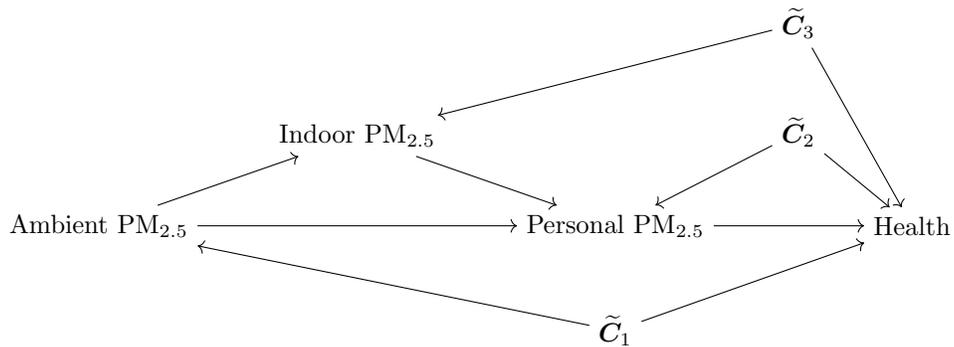
\begin{figure}[H]
	\centering
	\begin{tikzpicture}
	\node[text centered] (useless) {};
	\node[above = 0.8 of  useless, text centered] (indoor) {Indoor PM$_{2.5}$};
	\node[right = 2 of  useless, text centered] (personal) {Personal PM$_{2.5}$};
	\node[right = 1 of  personal, text centered] (useless2) {};
	\node[above = 0.8 of  useless2, text centered] (conf2) {$\widetilde{\bm C}_2$};
	\node[above = 0.8 of  conf2, text centered] (conf3) {$\widetilde{\bm C}_3$};
	\node[right = 2 of personal, text centered] (health) {Health};
	\node[below = 0.8 of  personal, text centered] (conf1) {$\widetilde{\bm C}_1$};
	\node[left = 2 of  useless, text centered] (pm) {Ambient PM$_{2.5}$};
	
	\draw[->] (pm) -- (personal);
	\draw[->] (indoor) -- (personal);
	\draw[->] (conf3) -- (indoor);
	\draw[->] (conf3) -- (health);
	\draw[->] (conf2) -- (personal);
	\draw[->] (conf2) -- (health);
	\draw[->] (pm) -- (indoor);
	\draw[->] (personal) -- (health);
	\draw[->] (conf1) -- (pm);
	\draw[->] (conf1) -- (health);
	\end{tikzpicture}
\caption{DAG relating exposure to ambient \PM{} concentrations to personal exposures and health outcomes. Covariate sets $\widetilde{\bm C}_1$, $\widetilde{\bm C}_2$, $\widetilde{\bm C}_3$ represent potentially different sets of confounders between the health outcome and ambient, personal or indoor \PM{} concentrations or exposures. Arrows represent potential (but not necessarily present) relationships.}
\label{fig:outdoor_personal}
\end{figure}

In order to agree with existing causal inference literature for continuous treatments, we often refer to measurements $X_i$ as zip code $i$'s \textit{exposure}, and a range of ambient pollution concentration as an \textit{exposure level}. However, a zip code's measurement of ambient concentration $X_i$ might be substantially different from the personal exposure of an individual residing in that zip code. \cref{fig:outdoor_personal} shows a hypothetical DAG relating ambient and indoor pollution concentrations with individuals' personal exposures and health outcomes. Ambient pollution concentrations act on an individual's outcome only through the individuals' personal exposures.

In this paper, we focus on estimating the causal effects of ambient \PM{}  on cardiovascular health outcomes.
That is, potentially, the most interesting question from a policy perspective, since policy regulations (and the NAAQS) are set based on the knowledge for the effect of ambient concentrations.
The implications of using ambient concentrations instead of personal exposures to study the effect of pollution concentrations are discussed in \cref{sec:discussion}.

\subsection{Potential presence of local confounding in our study}

In the case of binary treatments, whether a covariate acts as a confounder is often evaluated by checking whether there exists significant imbalance in the covariate distribution of the treated and control groups. For continuous exposures, there is no direct counterpart to covariate balance since units are not separated into two groups. Instead, exploratory analyses for the presence of confounding are often based on covariates' strength in predicting the exposure through regression models \citep{Imai2004causal}. Then, a covariate's p-value in a model for the exposure is used to investigate whether it might be a confounding variable.

We use a related approach to illustrate the potential presence of local confounding in our data. We considered two subsets of zip codes: 1) zip codes with low ambient concentrations ($<$8$\mu g\slash m^3$; 817 observations); and 2) zip codes with high ambient concentrations ($>$11.5$\mu g\slash m^3$; 672 observations). Even though this definition of the low and high exposure levels is arbitrary for the purpose of our illustration, this choice ensures a similar number of observations and similar range of exposure values within the two levels. Within each exposure level \textit{separately}, we considered a linear regression of ambient pollution concentration on each covariate, and evaluated the covariate's predictive strength through its p-value.

\begin{figure}[!b]
\centering
\includegraphics[width=0.95\textwidth]{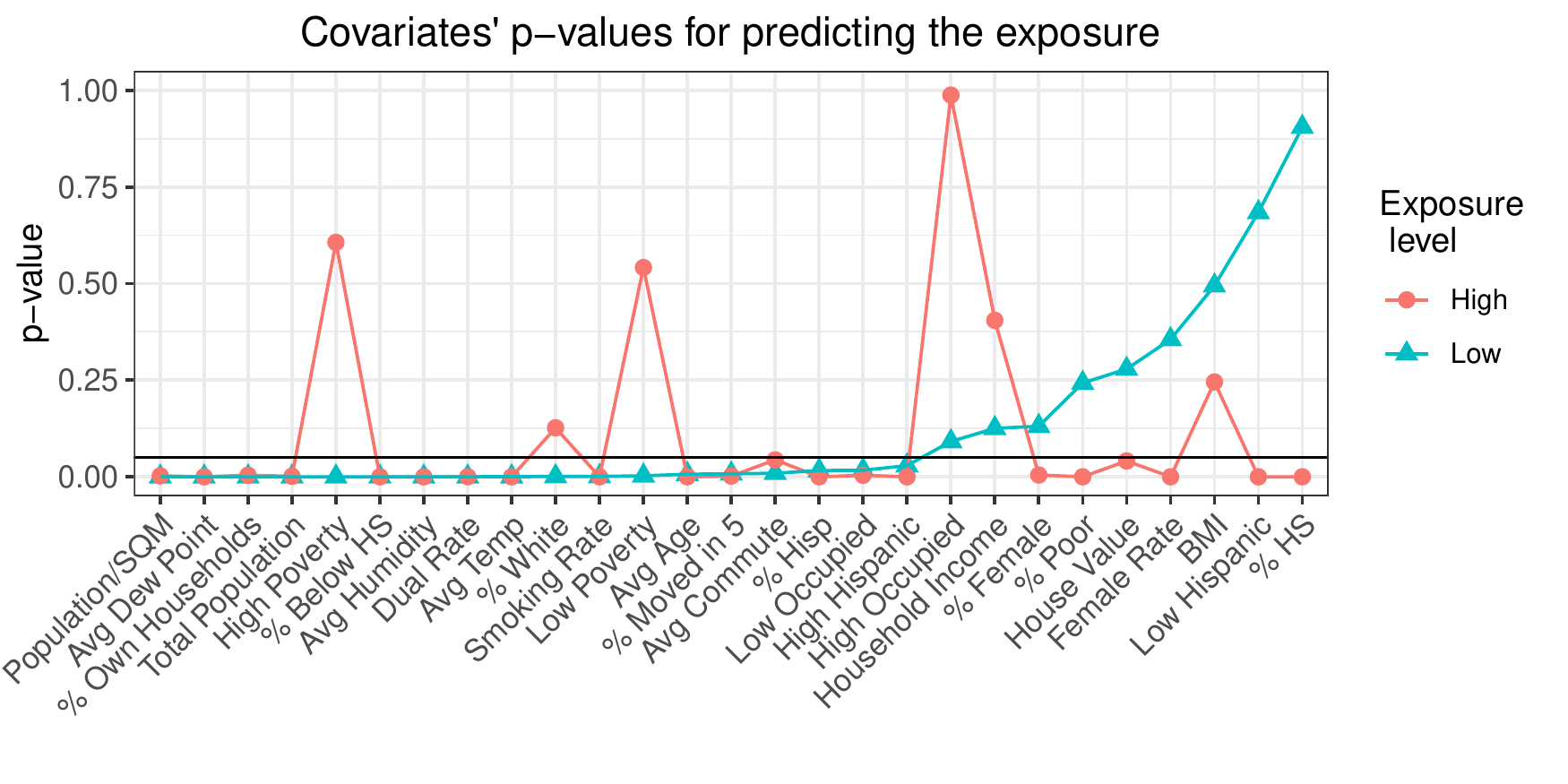}
\vspace{-20pt}
\caption{Covariate p-values in predicting the exposure, separately at the low (blue: $<8\mu g/m^3$) and high (red: $>11.5\mu g/m^3$) exposure levels.}
\label{fig:app_pvalues}
\end{figure}

\cref{fig:app_pvalues} shows the p-value of the covariates in the $p$ regression models, and for the two exposure levels.
We see that some variables such as population density (\texttt{Population/SQM}) and the percentage of the population with less than a high school education (\texttt{\% Below HS}) are predictive of ambient concentrations in both low and high exposure levels. However, other variables, such as the median household value (in logarithm -- \texttt{House Value}), are only predictive of ambient pollution concentrations at the high exposure levels. The opposite is true for the percentage of population that is white (\texttt{\% White}). 
Such initial investigation indicates that different variables might act as predictors of the ambient exposure at different exposure levels.

In Supplement \ref{app_sec:local_confounding}, we show the estimated covariates' coefficients whose p-values are shown in \cref{fig:app_pvalues}. Since the two exposure levels are relatively balanced in terms of number of observations and range of exposure values, the magnitude of the p-values in \cref{fig:app_pvalues} is directly comparable to the magnitude of the estimated coefficients. This indicates that initial investigation of local confounding could be equivalently performed in terms of estimated coefficients or p-values.

In Supplement \ref{app_sec:local_confounding}, we also consider a similar exploratory analysis to investigate which covariates are predictors of the health outcome at the low and high exposure levels separately. Combining the results presented there to the ones in \cref{fig:app_pvalues}, there is evidence that the variables that confound the ER relationship might differ across levels of the exposure leading to {\it local confounding}. For example, the zip code median household value (\texttt{House Value}) is predictive of both ambient air pollution and cardiovascular hospitalization rates at the high exposure levels, but is not predictive of ambient air pollution at low  exposure levels. Additionally, there is indication that the percentage of the population with less than a high school degree (\texttt{\% Below HS}) is a confounder at the low exposure levels, whereas the same variable is not predictive of the health outcome at the high exposure levels.


\section{Causal ER, the experiment configuration, and the local ignorability assumption}
\label{sec:notation}
We follow the potential outcome framework \citep{Neyman1923,Rubin1974,Hirano2004}, and under the stable unit value of treatment assumptions (SUTVA; no interference, no hidden versions of the treatment \citep{Rubin1980}), we use $Y_i(x)$ to denote the potential outcome for observation $i$ at exposure $x \in \mathcal{X}$, where $\mathcal X \subset \mathbb{R}$ is the interval including all possible exposure values.  Then, $\{Y_i(x), x \in \mathcal{X} \}$ is unit $i$'s ER curve, and  $\{ \overline Y(x) = E[Y_i(x)], x \in \mathcal{X}\}$ is the population average ER curve.

Assuming $\overline Y(x)$ is differentiable as a function of $x$, we define the instantaneous causal effect
\[
\Delta(x) = \lim_{h \rightarrow 0} \frac{\overline Y(x+ h) - \overline Y(x)} h.
\]
A  $\Delta(x) \neq 0$ implies that variation in the exposure in a neighborhood of $x$ has a causal effect on the expected outcome. We also define the population average causal effect of an exposure shift from $x$ to $x + \delta$, as $CE_\delta(x) = \overline Y(x + \delta) - \overline Y(x) = \int_x^{x + \delta} \Delta(t) \mathrm{d}t$.
The observed outcome $Y_i$ is equal to the potential outcome at the observed exposure $Y_i(X_i)$.

Under the weak ignorability assumption which states that the treatment is as if randomized conditional on observed covariates,
$X \independent Y(x) | \mb C$,
and every subject in the population can experience any $x \in \mathcal{X}$, $\overline Y(x)$ is identifiable using the observed data \citep{Hirano2004}.
Then, a minimal confounding adjustment set $\mb C^* \subseteq \mb C$ is a set of covariates which satisfies $X \independent Y(x) | \mb C^*$, but $X \notindependent Y(x) | \mb C^{**}$ for any $\mb C^{**}$ strict subset of $\mb C^*$ \citep{Luna2011a,Wang2012,Vansteelandt2012}.

\begin{figure}[H]
\begin{center}
\includegraphics[width = 0.65\textwidth]{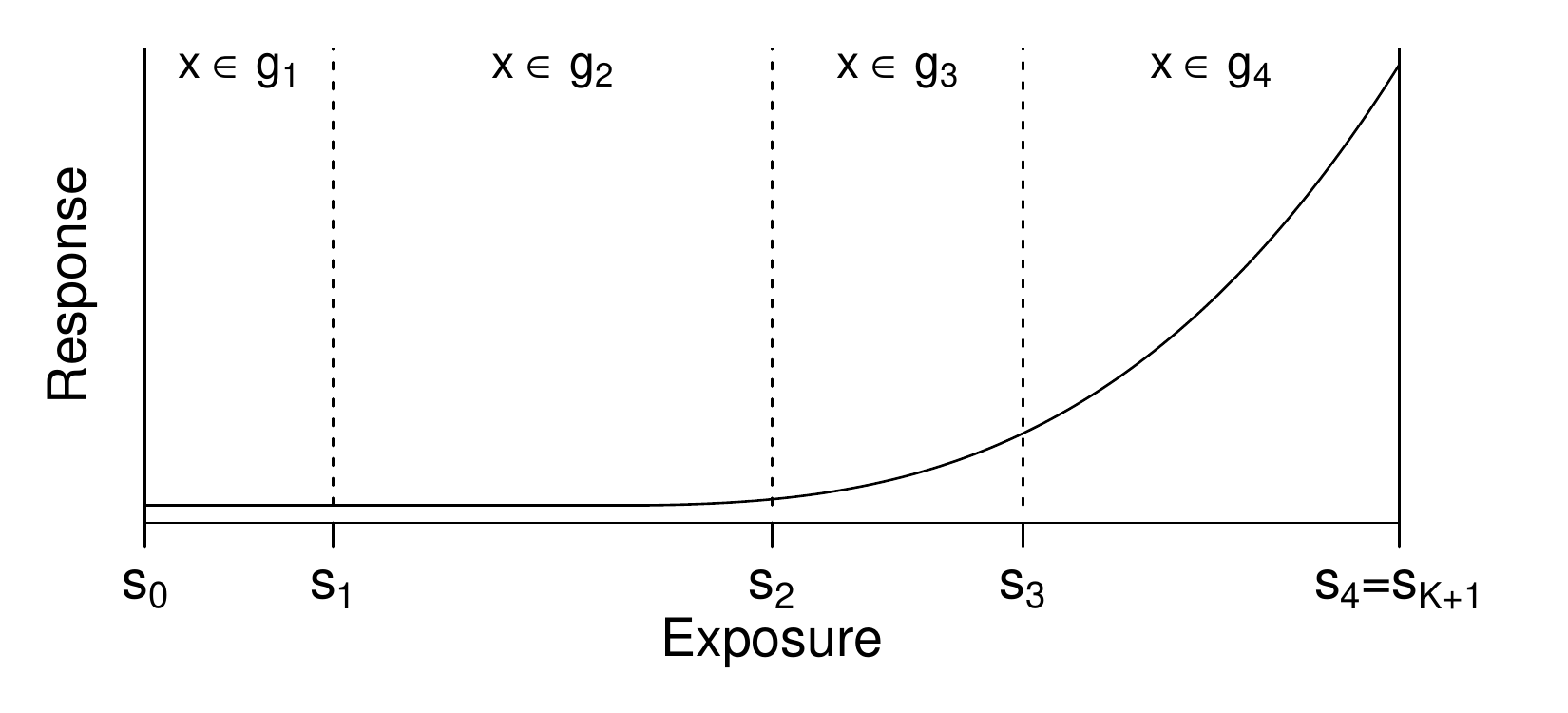}
\end{center}
\caption{ER curve with exposure range partitioned by $\bar{\mb s}$ in 4 experiments.}
\label{fig:ER_example}
\end{figure}

In this paper, we are interested in addressing the possibility that the minimal sufficient adjustment set $\mathbf C^*$ varies across exposure levels. We formalize this by introducing the \textit{experiment configuration}.
Let $K$ denote a fixed positive integer, and $\text{min}= \inf \mathcal X$ and $\text{max}=\sup \mathcal{X}$ denote the known and fixed minimum and maximum values of the exposure range $\mathcal{X}$. Then, $\mathbf{\bar{s}} = (s_0 = \min, s_1, s_2, \dots, $ $s_K, s_{K + 1} = \max)$ is the experiment configuration which defines a partition of the exposure range in $K + 1$ experiments $g_k = [s_{k-1}, s_k)$. We use $\mb s$ to denote the internal points $s_1, s_2, \dots, s_K$. In \cref{fig:ER_example}, a hypothetical exposure response function is plotted where $\mathbf{\bar{s}}$ defines a total of 4 experiments ($K = 3$).
Then, 
$\mb C^*_k$ is a minimal sufficient adjustment set in experiment $k$ if it satisfies
\begin{equation}
X \independent Y(x) | \mb C^*_k,\ \text{for all } x \in g_k,
\label{eq:ignor_exper}
\end{equation}
and (\ref{eq:ignor_exper}) does not hold for any strict subset of $\mb C^*_k$.
The sets $\mb C^*_k$ can be disjoint, identical, or overlapping if the same variable is necessary for confounding adjustment in more than one experiment.

\section{ER estimation in the presence of local confounding}
\label{sec:method}

Motivated by the evidence of local confounding between ambient \PM{} concentrations and cardiovascular hospitalizations discussed in \cref{sec:data_description}, we introduce 
LERCA: Local Exposure Response Confounding Adjustment. In order to build intuition, we do so for a fixed experiment configuration in \cref{sec:fixed_s}. LERCA with unknown $\mb s$ is presented in \cref{sec:unknown_s}. The choice of $K$ is discussed in \cref{subsec:choosing_K}.

\subsection{Known experiment configuration}
\label{sec:fixed_s}

Assume for now a known experiment configuration $\mathbf{\bar{s}}$. Then, locally, that is for $x \in g_k = [s_{k-1}, s_k)$, we assume the following pair of exposure and outcome models:
\begin{equation}{\small
    \begin{aligned}
    p(x |\mb C = \mb c, x \in g_k) & =
    \phi \big(x;\ \delta_{k0}^X + \textstyle{\sum_{j = 1}^p} \alpha_{kj}^X \delta_{kj}^X c_{j},\ \sigma_{k,X}^2 \big)\\
    p(y|X = x, \mb C = \mb c,  x \in g_k) & =
    \phi \big( y;\ \delta_{k0}^Y + \beta_k (x - s_{k - 1}) + \textstyle{\sum_{j = 1}^p} \alpha_{kj}^Y \delta_{kj}^Y c_j,\ \sigma_{k,Y}^2 \big)
    \end{aligned}}
\label{eq:likelihoods}
\end{equation}
where $\phi(\cdot;\mu, \sigma^2)$ denotes the normal density with mean $\mu$ and variance $\sigma^2$, and $\alpha_{kj}^X \in \{0, 1\}$ indicates that covariate $C_j$ is included into the exposure model of the $k^{th}$ experiment ($\alpha_{kj}^X=1$), or not ($\alpha_{kj}^X=0$). The parameter $\alpha_{kj}^Y$ has the same interpretation, but for the outcome model.
The parameter $\beta_k$ denotes the instantaneous change in the expected outcome associated with a local variation in exposure for $x \in g_k$, adjusted for the $C_j$s that have  $\alpha^Y_{kj}=1$.
Even though all parameters depend on which covariates are included in the corresponding model, we do not explicitly state this dependence for notational simplicity.
Model (\ref{eq:likelihoods}) allows for a different set of variables and variables' coefficients at the different experiments. 

If a minimal confounding adjustment set for experiment $k$ is included in the outcome model and the mean functional form is correctly specified, $\beta_k$ is an unbiased estimator of the instantaneous effect $\Delta(x)$, for $x \in g_k$. Similarly, an unbiased estimator of $\overline Y(x)$ is $E_{\bm C}\{ E[Y_i | X = x, \bm C] \}$, which can be estimated by taking the average over the units in our sample of the conditional expectation $E[Y_i | X = x, \bm C_i]$. 

In \cref{subsec:prior_alphas} we discuss how the prior distribution on the inclusion indicators is chosen to target confounding adjustment. In \cref{subsec:prior_continuity}, we discuss prior specification for outcome model coefficients that ensures borrowing of information across experiments and ER continuity across the exposure range. But first we address two questions that naturally arise from the specification of model \cref{eq:likelihoods}. First, we clarify the connection between LERCA and a model that specifies the ER relationship using linear splines in \cref{subsec:connection_splines}. Then, in \cref{subsec:connection_separate}, we discuss how LERCA compares to a model that is fit separately within each experiment $g_k$.

\subsubsection{Connection to linear splines}
\label{subsec:connection_splines}

In the outcome specification of model \cref{eq:likelihoods}, the term $\beta_k (x - s_{k-1})$ in the mean functional could be substituted by $\beta_k x$ and $- \beta_k s_{k - 1}$ could be absorbed in the intercept. However, specifying the model as to include $\beta_k (x - s_{k-1})$ demonstrates the connection between model \cref{eq:likelihoods} and a model where the ER relationship is specified using linear splines with knots $\bm s$. Furthermore, such specification significantly simplifies prior elicitation to ensure ER continuity (see \cref{subsec:prior_continuity}), and posterior sampling satisfying the continuity condition (see Supplement \ref{app_sec:MCMC}).

Even though the outcome model in \cref{eq:likelihoods} resembles a linear splines model, there is a \textit{key} distinction between the two models.
In model \cref{eq:likelihoods}, different experiments $g_k$ are allowed to have a different slope for the exposure ($\beta_k$), a different set of outcome predictors (covariates with $\alpha_{kj}^Y = 1$), or the same set of predictors but with different coefficient ($\delta_{kj}^Y$). Therefore, points $\mb s$ in \cref{eq:likelihoods} represent a change in the slope or a change in the outcome model covariate adjustment.
On the other hand, a model that uses splines for the exposure-response relationship only allows $\beta_k$ to vary with $k$. In this sense, a splines model is a sub-case of model \cref{eq:likelihoods}, that for $\alpha_{kj}^Y$ and $\delta_{kj}^Y$ constant across $k$.

The assumption of local linearity (linear effect of the exposure on the outcome within each experiment) can lead to global non-linearity, and can be relaxed using higher order splines. However, for our study of the health effects of ambient air pollution at low concentrations, previous research indicates that the relationship between air pollution and cardiovascular outcomes is linear \citep{Thurston2016ambient, Lim2018association} or supra-linear \citep{Crouse2015, Pinault2017associations}, situations that our model can adjust to.

\subsubsection{Connection to a separate model across experiments}
\label{subsec:connection_separate}

A natural question that arises from the LERCA model specification in \cref{eq:likelihoods}, is how LERCA compares to fitting a separate outcome model within each experiment $g_k$. Doing so would still allow for different confounders and different confounding strength at different exposure levels.

However, a separate model within each experiment would not borrow any information across exposure levels, and could estimate an ER that is not continuous at the points $\mb s$.
In contrast, LERCA borrows information across exposure levels by ensuring that the estimated ER is continuous everywhere (see \cref{subsec:prior_continuity}). If higher order polynomials are used within each experiment, LERCA, similarly to splines, could be altered to accommodate higher order smoothness across the exposure range.

\subsubsection{Prior distribution on inclusion indicators for confounding adjustment}
\label{subsec:prior_alphas}

We build upon the work by \cite{Wang2012, Wang2015} to assign an informative prior  on covariates' local inclusion indicators $(\alpha^X_{kj}, \alpha^Y_{kj})$. This prior choice ensures that model averaging assigns high posterior weights to outcome models including a minimal confounding adjustment set separately for each exposure range, and specifies
\begin{align*}
& \frac{P(\alpha^Y_{kj} = 1 | \alpha^X_{kj} = 1)}{P(\alpha^Y_{kj} = 0 | \alpha^X_{kj} = 1)} = \omega\; \mbox{where} \; \omega > 1,\; \mbox{iid} \; \forall \; j,k.
\numberthis
\label{eq:BAC_prior}
\end{align*}
By specifying \cref{eq:BAC_prior}, a variable $C_j$ is assigned high prior probability to be included into the outcome model if it is also included in the exposure model ($x \in g_k$ \& $\alpha^X_{kj} = 1$). \citet{Wang2012} and \cite{Antonelli2017guided} show that, for binary treatments, this informative prior leads to outcome models that include the minimal set of true confounders with higher posterior weights than model selection approaches that are based solely on the outcome model. In our context, this experiment-specific prior specification ensures that, locally, covariates in the minimal set $\mb C^*_k$ are included in the outcome model of experiment $k$ with high posterior probability.

\subsubsection{Ensuring ER continuity}
\label{subsec:prior_continuity}

In most applications, it is expected that the causal ER relationship $\overline Y(x)$ is continuous in $x$. Therefore, estimates of $\overline Y(x)$, in our case $E_{\bm C}\{ E[Y | X = x, \bm C]\}$, should also be continuous. If the covariates $C_j$ are centered, and under model \cref{eq:likelihoods}, continuity of the estimated ER function is satisfied if
\begin{align*}
& \lim_{x \rightarrow s_k^+} E_{\bm C} \{ E[Y | X = x, \bm C] \} = \lim_{x \rightarrow s_k^-} E_{\bm C} \{ E[Y | X = x, \bm C] \} \\
& \hspace{20pt} \iff
\delta_{k0}^Y = \delta_{(k-1)0}^Y + \beta_{k - 1}(s_k - s_{k - 1}).
\numberthis
\label{eq:intercept_prior}
\end{align*}
This is ensured by assuming a point-mass recursive prior on $\delta_{k0}^Y, k \geq 2$.
Then, conditional on $\mb s$, the outcome model intercept of experiment $k \geq 2$ is a deterministic function of the outcome model intercept of the first experiment $\delta_{10}^Y$, and the slopes $\beta_1, \beta_2, \dots, \beta_{k - 1}$. These parameters are assigned independent non-informative normal prior distributions.

\subsubsection{Prior distributions of the remaining coefficients}

Prior distributions on the remaining regression coefficients (exposure model coefficients, outcome model covariates' coefficients) and variance terms are chosen such that they lead to known forms of the full conditional posterior distributions to simplify sampling.
We use independent non-informative inverse gamma prior distributions on $\sigma^2_{k, X}, \sigma^2_{k, Y}$. 
Non-informative normal prior is chosen for the exposure model intercepts $\delta_{k0}^X$.
Conditional on the inclusion indicators, the prior on the regression coefficient $\delta_{kj}^Y$ is a point mass at 0, or a non-informative normal distribution when $\alpha_{kj}^Y$ is equal to 0 or 1 accordingly. Similarly for the exposure model covariates' coefficients $\delta_{kj}^X$.
Default hyperparameter values are set to 0.001 for the inverse gamma distribution, and (0, 100) for the mean, and standard deviation of the normal distribution.
Details on the prior specifications can be found in Supplement \ref{app_sec:priors}.

\subsection{Unknown experiment configuration}
\label{sec:unknown_s}

For a fixed experiment configuration $\mathbf{\bar s}$, each experiment is treated separately in terms of confounder \textit{selection and strength} of the confounding adjustment. However, the configuration itself is a key component of the fitted exposure response curve, and fixing it a priori could lead to bias and uncertainty underestimation. Instead, we assume that, a priori, the internal points of the experiment configuration $\mb{s}$ are distributed as the even-numbered order statistics of $2K + 1$ samples from a uniform distribution on the interval $(s_0, s_{K + 1})$.
This prior choice of $\mb s$ discourages specifications of $\mb s$ that include values that are too close to each other \citep{Green1995}.
The prior is augmented by indicators that consecutive points $s_k, s_{k+1}$ cannot be closer than some distance $d_k$.
Conditional on $\mb s$, we follow the model specification and prior distributions described in \cref{sec:fixed_s}.

\subsection{MCMC scheme and convergence diagnostics}

Markov Chain Monte Carlo (MCMC) methods are used to acquire samples from the posterior distribution of model parameters. A detailed description of the MCMC scheme including computational challenges and contributions can be found in Supplement \ref{app_sec:MCMC}. There, we also discuss MCMC convergence diagnostics based on the potential scale reduction factor (PSR; \cite{Gelman1992}) for quantities that do not directly depend on the experiment configuration.

\subsection{Number of points in the experiment configuration}
\label{subsec:choosing_K}

As presented previously, LERCA requires the specification of the internal number of points $K$ in the experiment configuration. Since the number of parameters grows with $K$, possible values for $K$ could be bounded by considering the maximum number of coefficients we are willing to entertain.

Cross validation methods to choose values of tuning parameters are often infeasible in the Bayesian framework due to time and computational resources constraints.
In a comprehensive review, \cite{Gelman2014} discusses methods of estimating the expected out of sample prediction error for Bayesian methods.
The widely-applicable information criterion (WAIC; \cite{Watanabe2010}) provides an estimate of the out-of-sample prediction error based on one MCMC run. It is defined as $WAIC = - 2 \left(\mbox{lppd}- p_{WAIC} \right)$, 
where $\mbox{lppd}$ and $p_{WAIC}$ denote the 
log point-wise posterior predictive density and the penalty:
 \begin{align*}
\mbox{lppd} = &\sum_{i = 1}^n \log E_{post}p(x_i, y_i | \theta)\\
p_{WAIC} = & \sum_{i = 1}^n \mathrm{var}_{post} \left( \log p(x_i, y_i | \theta) \right).
\end{align*}
Here, $\theta$ denotes the full vector of parameters, and $E_{post}$, $\mathrm{var}_{post}$ denote the posterior mean and variance.

In order to choose $K$ for LERCA, LERCA is fit \textit{once} for different values of $K$, and $K$ is chosen as the value that minimizes the estimate of the WAIC. 


\section{Simulation Studies}
\label{sec:simulations}

The main goal of our simulation study is to illustrate that local confounding is an important issue that both commonly-used and flexible approaches for ER estimation fail to adjust for and they return biased results. The results from our simulation study indicate that methodology that directly accommodates local confounding is necessary in order to correctly estimate the causal effect of a continuous exposure. An R package which can be used to generate data with local confounding and fit LERCA is available at \url{https://github.com/gpapadog/LERCA}.

Additionally, in \cref{sec:sims_global} we discuss results from a simulation study under a generative model \textit{without} local confounding. In this case, traditional approaches and global confounding adjustment suffice for ER estimation, and the question is how comparably LERCA performs.

The approaches we considered are:
\begin{enumerate}
	\item Generalized Additive Model (\texttt{GAM}): Regressing the outcome $Y$ on flexible functions of the exposure $X$ and all potential confounders (4 degrees of freedom for each predictor).
    \item Spline Model (\texttt{SPLINE}): Additive spline estimator described in \cite{Bia2014}. The generalized propensity score (gps) is modelled as a linear regression on all covariates. The ER function is estimated using additive spline bases of the exposure and gps.
    \item The Hirano and Imbens estimator \citep{Hirano2004} (\texttt{HI-GPS}):
ER estimation is obtained by fitting  an outcome regression model including quadratic terms for both the exposure and the gps, and the exposure-gps interaction. The gps is estimated as in \texttt{SPLINE}.
    \item Inverse Probability Weighting estimator (\texttt{IPW}): The generalized propensity score is used to weigh observations in an outcome regression model that includes linear and quadratic terms of exposure. The gps is estimated as in \texttt{SPLINE}.
    \item The doubly-robust approach of \cite{Kennedy2017} (\texttt{KENNEDY}): The gps and outcome models are estimated using the Super Learner algorithm \citep{vanderlaan2007} combining the sample mean, linear regression with and without two-way interactions, generalized additive models, multivariate adaptive regression splines, and random forests.
Based on the gps and outcome model estimates, the pseudo-outcome is calculated and is regressed on the exposure using kernel smoothing. This approach is chosen to represent state-of-the-art methods in ER estimation that are based on flexible, machine-learning and non-parametric approaches.
\end{enumerate}

\subsection{Data generation with local confounding}
We generate data with exposure values which range from 0 to 10 and are uniformly distributed over the exposure range. Even though a uniform distribution is not accurate for the exposure variable in our study (ambient air pollution concentrations), we consider a uniformly distributed exposure to ensure that methods' performance is solely affected by the presence of local confounding, and not by the presence of limited sample size at some exposure levels. We consider a quadratic ER, and true experiment configuration $\mb{\bar{s}} = (0, 2, 4, 7, 10)$. Table \ref{tab:sim_cov} summarizes which of the 8 potential confounders are predictive of the exposure and$\slash$or the outcome within each experiment (correlations and regression coefficients are summarized in \cref{app_table:sims_covs}). Note that in this data generating mechanism the minimal set of confounders vary across the four experiments. We simulate 400 data sets of 800 observations each. Details on the data generating mechanism are in Supplement \ref{app_sec:sim_mech}. 

\begin{table}[H]
\centering
\caption{Representation of which covariates are predictive of the exposure and $\slash$ or the outcome within each experiment (denoted by a \checkmark). Covariates with \checkmark in both models within the same experiment are  local confounders.}
\label{tab:sim_cov}
\begin{tabular}{cr|cccccccc}
Experiment & Model & $C_1$ & $C_2$ & $C_3$ & $C_4$ & $C_5$ & $C_6$ & $C_7$ & $C_8$  \\ \hline
1 & $X | \mb C$ & \checkmark & \checkmark & \checkmark &&&& \\
  & $Y|X, \mb C$ & \checkmark & \checkmark & \checkmark & & & & & \\ \hline
2 & $X | \mb C$ & \checkmark & \checkmark & & \checkmark & & & & \\
  & $Y | X, \mb C$ & & \checkmark & \checkmark & \checkmark & & & & \\ \hline
3 & $X | \mb C$ &\checkmark & & \checkmark & & \checkmark & & &  \\
  & $Y | X, \mb C$ & & \checkmark & \checkmark & & \checkmark &&& \\ \hline
4 & $X | \mb C$ & & \checkmark & & & \checkmark & \checkmark & & \\
  & $Y | X, \mb C$ & & \checkmark & \checkmark & & & \checkmark && \\ \hline
\end{tabular}
\end{table}

\subsection{Fitting the methods}

The different methods are fit using the \texttt{gam} and \texttt{causaldrf} R packages \citep{Hastie2017,Schafer2015}, and the code available on \cite{Kennedy2017}. LERCA is fit for $K \in \{2, 3, 4\}$, and for each data set the results shown correspond to the $K$ that minimized the WAIC.

Using each method, we estimate the population average ER curve $\overline Y(x)$ over an equally spaced grid of points on the interval $(0, 10)$, and compare the root mean squared error (rMSE) as a function of $x$. We also assess whether LERCA can recover the correct location of the points $\bm s$, identify the true confounders within each experiment, and choose the correct value for $K$.

\subsection{Simulation Results}

\begin{figure}[!t]
\centering
\includegraphics[width = 0.98\textwidth]{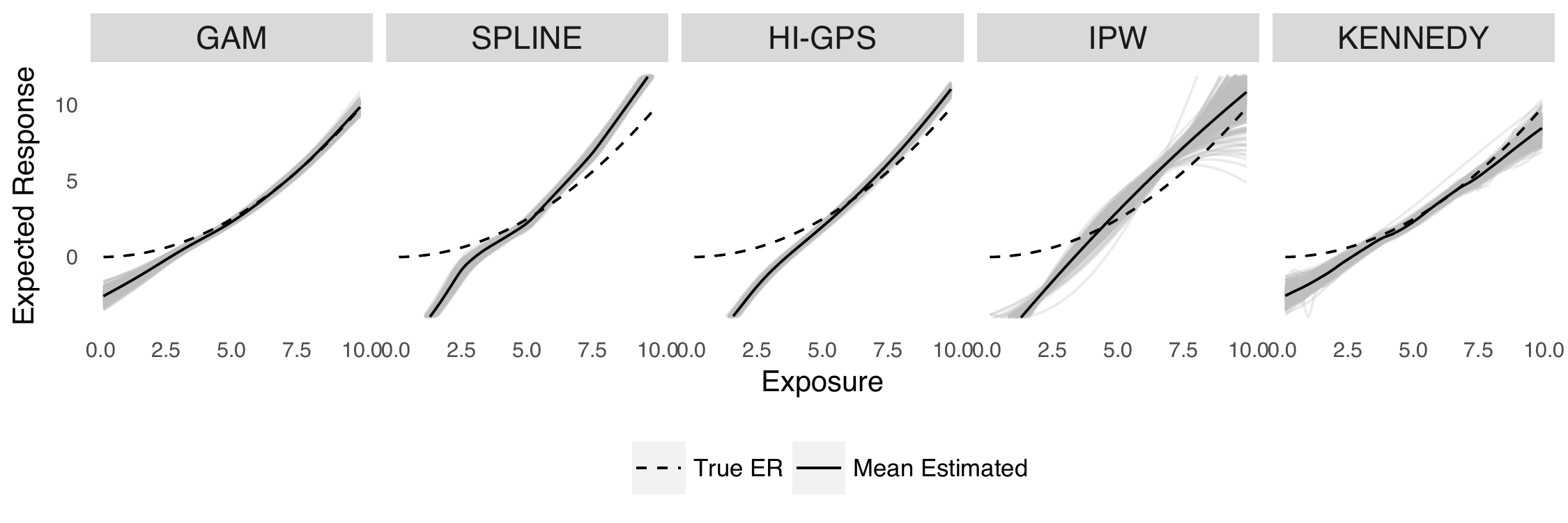}
\caption{The true mean ER function (dashed line), estimated ER functions from each simulated data set (gray), and the mean of the estimated ER functions (solid lines) using all alternative methods.}
\label{fig:sims_others}
\vspace{10pt}
\includegraphics[width = 0.98 \textwidth]{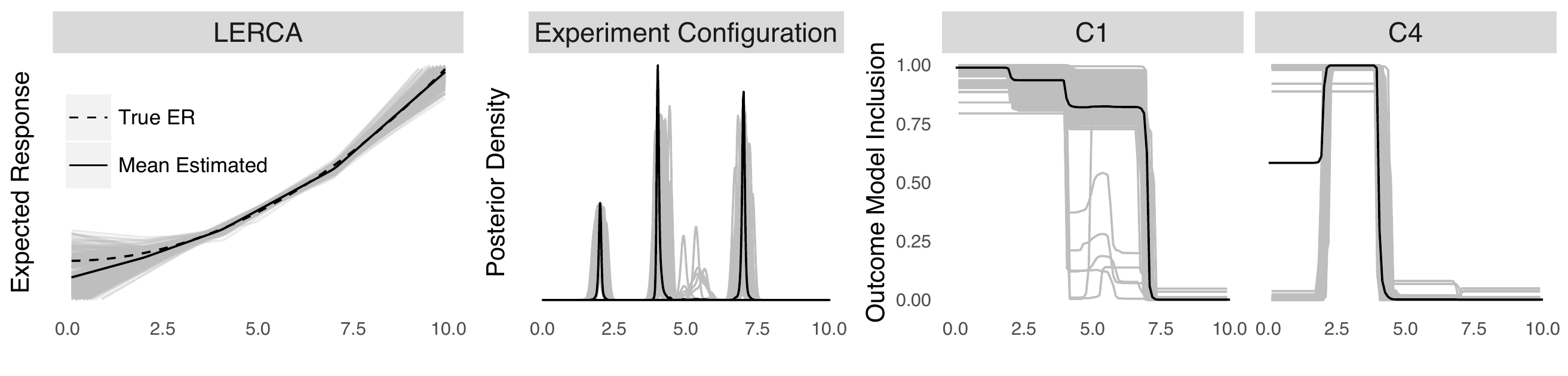}
\caption{LERCA results. (Left) Mean ER estimates. (Center) Posterior distribution of the internal locations $\mb s$. (Right) Outcome model posterior inclusion probability of $C_1$ and $C_4$. Gray lines correspond to simulated data sets separately, and black solid lines correspond to averages across data sets.}
\label{fig:sims_LERCA}
\end{figure}

\cref{fig:sims_others} shows the estimated ER curves using the alternative methods.
In \cref{fig:sims_LERCA} we summarize the LERCA results including the estimated ER, the internal points of the experiment configuration and outcome model inclusion indicators of covariates $C_1, C_4$ as a function of exposure $x \in (0, 10)$. We choose $C_1$ and $C_4$ because, in this data generating mechanism, $C_1$ is a confounder in experiment 1 ($x < 2$), and $C_4$ is a confounder in experiment 2 only ($2 < x < 4$).
Grey lines correspond to results for individual data sets, whereas black solid lines correspond to averages across simulated data sets.

In \cref{fig:sims_others} we see that the alternative methods return biased results, especially at very low or very high levels of the exposure. These results indicate that neither commonly-used nor flexible approaches utilizing machine learning tools appropriately accommodate local confounding adjustment for ER estimation.
In terms of root MSE, LERCA was consistently lower than the alternative methods at low exposure levels, all approaches performed similarly at middle exposure levels, and GAM slightly outperformed LERCA at high levels (\cref{fig:sims_rootMSE}). In Supplement \ref{app_subsec:sims_local_reverse}, we show that the relative performance of GAM and LERCA is reversed when the confounding structure is also reversed. These results indicate that local confounding is an issue across all exposure levels, and that, since the true confounding structure is never known for a real data set, LERCA should be preferred if local confounding is of concern.

As showed in \cref{fig:sims_LERCA}, even though the true ER is quadratic and LERCA is formulated as piece-wise linear, LERCA is able to identify the correct shape of the exposure-response function.  We find that using WAIC to choose the value of $K$ led to choosing the correct value of $K = 3$ 40\% of the times, and $K = 2$ 58\% of the times indicating that WAIC tends to over-penalize large values of $K$. Regardless, the correct internal locations $\mb s = \{ 2, 4, 7\}$ are located at the modes of the posterior distribution (second panel in \cref{fig:sims_LERCA}).
By examining the posterior inclusion probabilities of $C_1, C_4$, we observe that instrumental variables (e.g., $C_1$ in experiments 2 and 3) are often included in the outcome model. However, LERCA includes the minimal confounding set within each experiment with very high probability. On average (across the points in the exposure range and across all the simulated data sets) the minimal confounding set was included in the adjustment set 99\% of the times (ranging from 89-100\% across simulated data sets), indicating that the variables necessary for confounding adjustment are almost always included in the adjustment set.
Lastly, the point-wise 95\% and 50\% credible intervals cover the true mean ER values 84\% and 39\% of the times accordingly. The observed under-coverage is largely due to the underestimation of $K$.

\subsection{Simulation results in the absence of local confounding}
\label{sec:sims_global}

The previous generative scenario compared methods' performance in the presence of local confounding. In Supplement \ref{app_subsec:sims_global}, LERCA is compared to the alternative methods in the more traditional setting of global confounding, that is, in the setting more favorable to the other methods. In this context, LERCA with $K = 3$ (fixed) performed similarly in terms of root MSE compared to GAM and Kennedy's doubly-robust estimator, but better than the remaining alternative methods. These results indicate that LERCA offers a protection against bias arising from local confounding, without sacrificing efficiency when local confounding is not present.
 
\section{Estimating the effect of ambient \PM{} concentrations on zip code cardiovascular hospitalization rates}
\label{sec:application}

We estimate the relationship between the average ambient \PM{} concentrations for the years 2011-2012 and log cardiovascular hospitalization rates in 2013, using the data set introduced in \cref{sec:data_description}, and allowing for local confounding adjustment. Here, a unit $i$ from \cref{sec:notation} corresponds to the areal unit of a zip code.

\subsection{Plausibility of the causal assumptions in our study}
\label{subsec:plausibility}

The interpretation of estimated results as causal are bound by the plausibility of the causal assumptions within the study's setting. Here, we examine these assumptions in the evaluation of the causal relationship between ambient \PM{} and cardiovascular hospitalization rates.

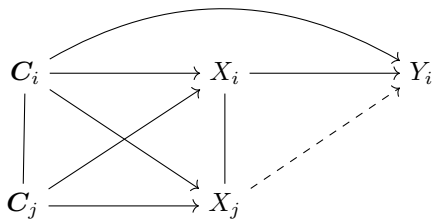
\begin{figure}[H]
	\centering
	\begin{tikzpicture}
	\node[text centered] (X1) {$X_i$};
	\node[left = 2 of  X1, text centered] (C1) {$\bm C_i$};
	\node[below = 1.2 of  X1, text centered] (X2) {$X_j$};
	\node[left = 2 of  X2, text centered] (C2) {$\bm C_j$};
	\node[right = 2 of X1, text centered] (Y1) {$Y_i$};

	\draw[->] (C1) -- (X1);
	\draw[->] (C1) to [out=30,in=150,looseness=1] (Y1);
	\draw[->] (C2) -- (X2);
	\draw[->] (C1) -- (X2);
	\draw[->] (C2) -- (X1);
	\draw[-] (X1) -- (X2);
	\draw[-] (C1) -- (C2);
	\draw[->] (X1) -- (Y1);
	\draw[->, dashed] (X2) -- (Y1);
	
	\end{tikzpicture}
\caption{Subgraph relating zip code $i$ and $j$'s covariates and ambient \PM{} concentrations to $i$'s potential outcomes. Arrow from $X_j$ to $Y_i$ is weaker than that from $X_i$.}
\label{fig:assumption_plausibility}
\end{figure}

One assumption discussed in \cref{sec:notation} is SUTVA which states that a zip code's potential outcomes are only a function of the zip code's own \PM{} levels. If Medicare beneficiaries residing within a zip code travel outside of it, then other zip codes' ambient \PM{} concentrations can affect the personal exposures of zip code $i$'s beneficiaries and, as a consequence, the zip code's hospitalization rates, invalidating SUTVA. This phenomenon is referred to in the literature as \textit{interference}. Since \PM{} concentrations in nearby zip codes are similar, and Medicare beneficiaries are expected, at some level, to spend most of their time within a relatively close distance to their home, interference can be assumed to be limited (dashed arrow in \cref{fig:assumption_plausibility}). When interference is limited, \cite{Savje2018average} showed that ignoring it returns estimates that are close to an average treatment effect.

The most commonly invoked causal assumption is that of ignorability. In our setting, ignorability implies that, conditional on measured covariates, a zip code's ``assignment'' to a specific level of ambient \PM{} does not depend on its \textit{potential outcomes} (see \cref{eq:ignor_exper}), and any zip code can experience any ambient \PM{} within the observed range.
One natural question is whether spatial correlation of ambient pollution concentrations invalidate ignorability, either through confounding or positivity.
The no unmeasured confounding assumption is expected to hold if the set of measured covariates includes all confounders.
In our study, we might expect that the covariates of nearby zip codes (such as nearby weather conditions) affect a zip code's ambient \PM{} concentrations (arrow from $\bm C_j$ to $X_i$ in \cref{fig:assumption_plausibility}).
If, in addition, a zip code's outcome directly depends on the covariates of other zip codes (arrow from $\bm C_j$ to $Y_i$) then $\bm C_j$ has to be included in the model for $Y_i$. We assume that such direct dependence does not exist in our study. For example, weather conditions near but not in zip code $i$ only affect zip code $i$'s hospitalization rates through their effect on ambient \PM{} concentrations.
Even though ambient \PM{} concentrations are spatially correlated, the positivity assumption requires that zip codes can experience any \PM{} concentration level \textit{marginally}, and the spatial correlation of \PM{} does not further complicate the plausibility of the positivity assumption.

Lastly, the interpretation of our study results as causal are bound by the specification of the model in \cref{eq:likelihoods}. If the mean functionals are not correctly specified, estimates of $\beta_k$ can be biased for the causal effect of \PM{} within that exposure level. Even though model \cref{eq:likelihoods} assumes independence of \PM{} concentrations, the spatial dependence structure is not expected to affect estimation of the model's regression coefficients or variable selection.

The results presented next can only be interpreted as causal under the assumptions discussed here. If any of the assumptions is violated, the study results should be interpreted as associational.

\subsection{Study results}
\label{subsec:study_results}

\begin{figure}[!t]
\centering
\includegraphics[width =  0.98\textwidth]{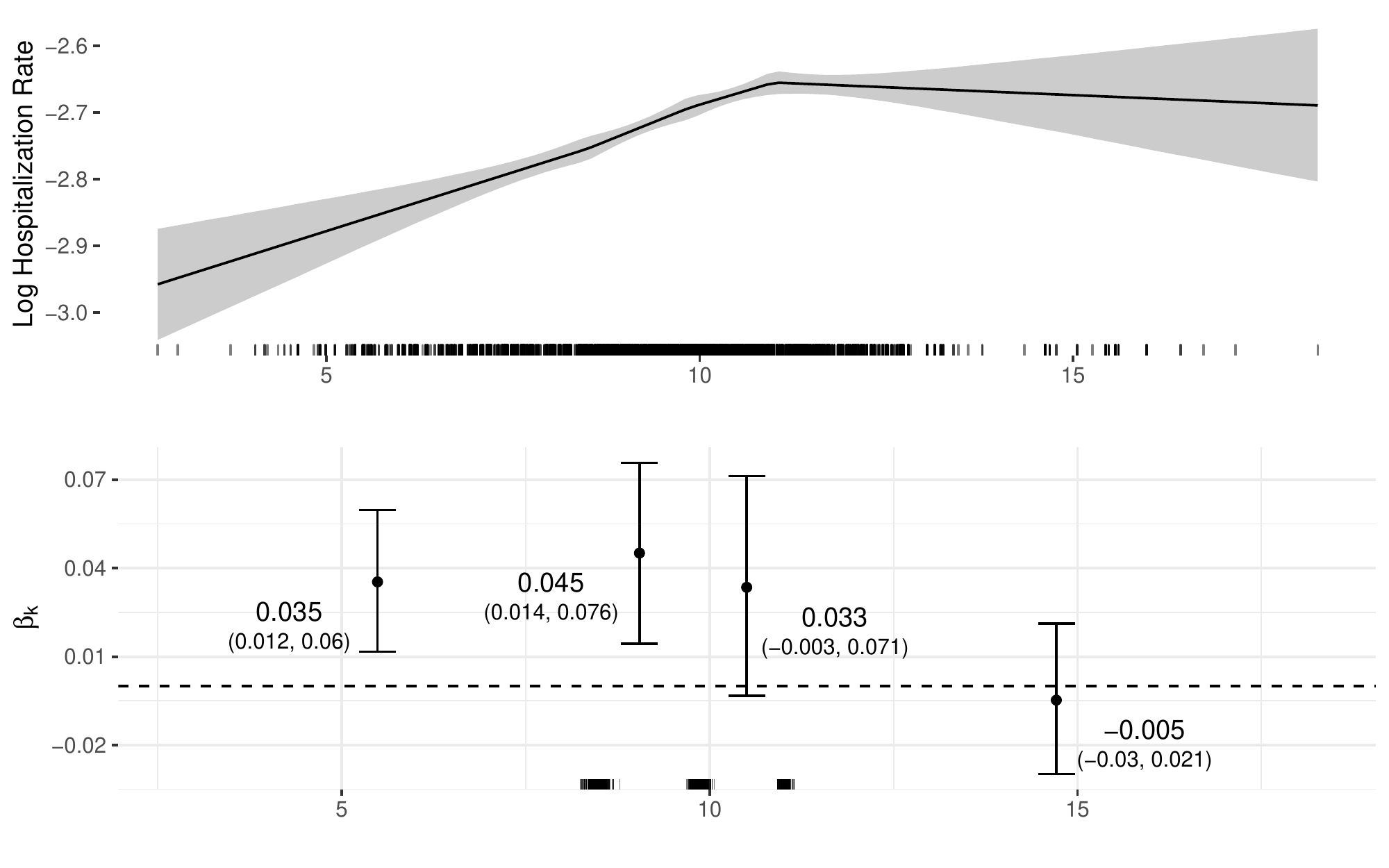}
\caption{ Top: Mean ER curve of PM$_{2.5}$ exposure (x-axis) and log all-cause cardiovascular hospitalizations (y-axis) --solid line-- with 95\% pointwise credible intervals. The rug of points shows the distribution of observed PM$_{2.5}$ values. Bottom: The posterior mean and 95\% credible interval of the $\beta$ coefficient within the four experiments. The rug of points shows the posterior distribution for $\mb s$ for $K = 3$.}
\label{fig:app_result}
\end{figure}

We fit LERCA for $K \in \{2, 3, \dots, 6\}$ and we report the results for $K=3$ which corresponds to the model with the lowest WAIC.
\cref{fig:app_result} shows the posterior mean and the 95\% credible intervals of the ER, the posterior distribution of the internal points of the experiment configuration, and the posterior mean and 95\% credible interval of $\beta_k$ within each experiment. Positive values of $\beta_k$ imply that an increase in ambient PM$_{2.5}$ concentrations leads to an increase in hospitalization rates.

In \cref{fig:app_result}, we see that the estimated ER is supra-linear with steeper incline at low concentrations. Examining the 95\% credible intervals for $\beta_k$, there is evidence that an increase in \PM{} at the low levels ($\leq 9.9\mu g/m^3$) leads to an increase in log hospitalization rates. However, 95\% credible intervals for $x \geq 9.9\mu g/m^3$ include zero.
Note that the current NAAQS for long term exposure to ambient PM$_{2.5}$ is equal to $12 \mu g\slash m^3$.
These results indicate that there is no exposure threshold for the effect of PM$_{2.5}$ on cardiovascular outcomes, which means that reductions in ambient PM$_{2.5}$ would lead to further health improvements, even at the low levels.
These results are consistent with other epidemiological studies which have found that the strength of the association between \PM{} and health outcomes is larger at low concentration levels \citep{Dominici2002, Shi2016,Di2017air}.
Lastly, the posterior distribution of $\mb s$, shows that observations below $8 \mu g / m^3$ and over $11.5 \mu g/m^3$ are always in the same experiment.

\subsection{Variability of the covariates' posterior inclusion across \PM{} concentration levels}

\begin{figure}[!t]
\includegraphics[width = \textwidth]{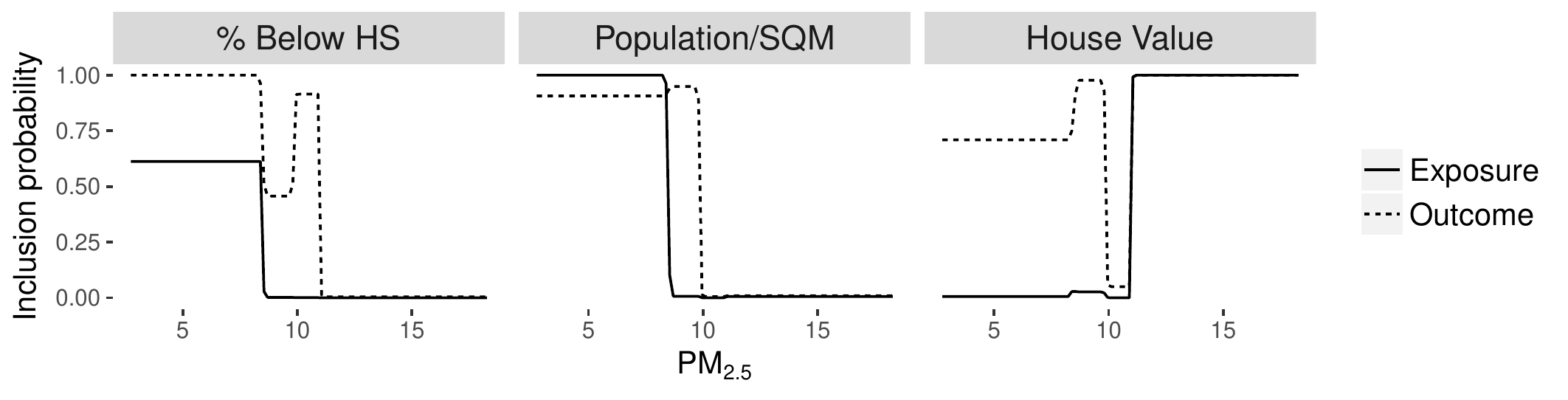}
\caption{Posterior inclusion probability of zip code population percentage with less than a high school education, population density, and median house value in the exposure and outcome model as a function of \PM{}.}
\label{fig:post_inclusion}
\end{figure}

We investigated whether local confounding was present by examining the variability of the covariates' inclusion probabilities in the exposure and outcome models as a function of \PM{}. \cref{fig:post_inclusion} shows the posterior inclusion probabilities for three covariates as a function of \PM{} providing a measure of the covariates' confounding importance across the \PM{} concentration range.

The posterior inclusion probabilities vary substantially at different concentration levels indicating that local confounding is likely to be present.
In \cref{fig:app_pvalues}, the exploratory analysis showed that the zip code median household value (\texttt{House Value}) was predictive of both \PM{} and hospitalization rates at the high ambient concentration levels, but only of the outcome at the low levels. The LERCA results in \cref{fig:post_inclusion} lead to the same conclusion. Similarly, the posterior inclusion probability for the variable representing the zip code's percentage of the population with less than a high school education (\texttt{\% Below HS}) indicates that this variable is an important confounder only at the low levels, in accordance to the exploratory analysis. LERCA returns a similar conclusion about the variable representing population density (\texttt{Population/SQM}), in disagreement with the analysis in \cref{sec:data_description} which showed that population density was predictive of both \PM{} concentrations and the outcome at both low and high levels.
Comparisons between the results in \cref{fig:app_pvalues} and the outcome model posterior inclusion probabilities were performed for all variables. LERCA tends to include in the outcome model a smaller number of variables than what one might have assumed based on the exploratory analysis. This is expected since LERCA considers the confounding importance of all variables simultaneously.

\subsection{Variability of the covariates' posterior inclusion within the low experiment}

With the focus of our study being the evaluation of the effect of ambient \PM{} at the low concentration levels, we studied the interpretation of $\beta_1$ across MCMC samples. Since the interpretation of $\beta_1$ as a causal effect requires that a sufficient adjustment set is included in the outcome model, we examined the variability of the covariates' outcome model inclusion indicators within the low experiment across iterations of the MCMC.

Across MCMC samples, 174 combinations of the covariates were included in the outcome model (out of the $2^{27}$ possible ones).
Even though this is a large number of potential models, 51\% of the posterior weight was given to the model with the following 9 covariates: the zip code's median house value and percentage of the population with at most a high school education, as measured in the 2000 Census and its extrapolation between 2000 and 2013, the population rate that has been a smoker at some point in their lives, the zip code's population density, the average dew point, the average age of Medicare beneficiaries and the percentage of them that are women. We refer to this model as \textit{Model 1}. The model with the second highest posterior probability included the same covariates except for smoking rate, and accounted for 7\% of the MCMC samples. Therefore, there is evidence that Model 1 outperformed the rest in confounding adjustment at low levels.

In order to evaluate the impact of model averaging on our final estimates, we compared the posterior distribution of $\beta_1$ across all MCMC samples to its distribution based on the MCMC samples for which Model 1 was chosen. Across all samples, $\beta_1$ was estimated to be equal to $0.035$ with 95\% credible interval $0.012-0.06$, and posterior probability that it is greater than 0 equal to 99.7\%. Among posterior samples for which Model 1 was chosen, $\beta_1$ was estimated to be $0.034$ with 95\% credible interval $0.011-0.056$ and posterior probability that $\beta_1$ is greater than 0 also equal to 99.7\%. The consistency of the Model 1 estimates and the model averaged estimates is an indication that model averaging, in this situation, did not lead to averaging over incompatible models.


\section{Discussion}
\label{sec:discussion}
We have introduced an innovative Bayesian approach for flexible estimation of the ER curve in observational studies that has the following features: 1) it casts the formulation of the ER within a potential outcome framework, and mimics several randomized experiments across exposure levels;
2) it uses the data to inform the experiment configuration; and given the current experiment configuration
3) allows for the possibility which is a reality in our study (\cref{fig:app_pvalues} and \cref{fig:post_inclusion}) that different sets of covariates are confounders at different exposure levels;
4) allows for varying confounding effect across levels of the exposure;
5) performs local covariate selection to increase efficiency;
6) propagates model uncertainty for the experiment configuration and covariate selection in the posterior inference on the whole ER curve;
and finally,
7) provides important scientific guidance related to which covariates are confounders at different exposure levels.

Although non-parametric and varying coefficient approaches \citep{Hastie1993} for ER estimation could, in theory, allow for differential confounding across different exposure levels, none of the existing methods for ER estimation explicitly accommodates local confounding, nor provides guidance for which covariates are confounders of the effect of interest at different levels of the exposure. Furthermore, the use of non-parametric methods to estimate a generalized propensity score or model the outcome of interest could prove unfruitful in situations where most of the available data are over a specific range of the exposure variable, the number of potential confounders is large, and interest lies in the estimation of causal effects for a change in the exposure in the tails of the exposure distribution.
In such situations, LERCA provides a way to model the outcome acknowledging that the exposure-response relationship might be confounded by different covariates at different exposure levels.
Lastly, it is worth noting that LERCA shall not be seen as a direct competitor to the approach by \cite{Kennedy2017}.
In fact, since the Super Learner algorithm combines different approaches for modeling the outcome, LERCA could be incorporated in the algorithm as an approach that allows for the presence of local confounding.

The main contribution of this paper is in addressing the issue of local confounding in ER estimation, and in providing guidance of covariates' confounding importance at different exposure levels. In doing so, LERCA is based on several modeling decisions that can be easily altered.
First, within each experiment and thus locally within a narrow exposure range, LERCA assumes linearity for both the outcome and exposure models. Local linearity could be relaxed by using higher order polynomials.
Second,  the informative prior on the inclusion indicators could lead to the inclusion of instrumental variables in the outcome model, which will not lead to bias, but will decrease the efficiency of our estimators. In the study of air pollution, strong instrumental variables are not expected to be present. Alternative strategies for local confounder selection can be accommodated here, extending, for example, work by \cite{Wilson2014,Cefalu2017} and \cite{Antonelli2018}. 
An interesting line of research is to explore LERCA's extensions to more flexible functional specifications, and to evaluate the performance of different approaches to model selection (via prior specifications or penalization techniques) for different confounding scenarios.

An alternative modification of LERCA could enforce that the ER curve is monotone, by assuming prior distributions on $\beta_k$ that are left (or right)-truncated at zero. This modification could be of explicit interest for environmental and toxicological research, and in studies of air pollution in particular where the ER relationship is often believed to be supra-linear \citep{Pinault2017associations, Vodonos2018concentration}. In risk assessment studies, the shape of the ER can greatly affect conclusions \citep{Pope2015health}, and is often assumed to be linear, log-linear, log-log, or power function with or without threshold \citep{Devos2016effect, Burnett2014integrated}. \cite{Nasari2016class} proposed a class of models that can capture various ER shapes and are easy to implement in large data sets. Even though LERCA can capture effectively any ER shape, development of a faster and computationally efficient estimation procedure is required for very large data sets. Future work could focus on incorporating local confounding adjustment in air pollution analyses including the whole United States and including zip codes that are not located near an air pollution monitor.

The results of our study are in agreement with a supra-linear ER shape, indicating that there is a larger health impact of ambient air pollution concentrations at low exposure levels, and that, if the causal assumptions hold, lowering ambient \PM{} concentrations would lead to a reduction in cardiovascular hospitalization rates.
Even though our analysis addresses a key question in air pollution epidemiology, it is also met with its own challenges.

First, even though focusing on ambient pollution concentrations is important from a policy perspective (since regulations can more directly control ambient concentrations), the results of this study are not directly interpretable for evaluating the health effects of \textit{personal} exposure to \PM{}. The relationship between ambient concentrations and personal exposures is complicated, with studies showing that there is substantial variability in personal exposures among individuals of similar ambient exposure concentrations \citep{Dockery1981personal, Clayton1993particle}, largely due to the individuals' activities \citep{Meng2009determinands}. 
The correlation between ambient concentrations and personal exposures might differ by exposure levels if, for example, individuals residing in highly polluted areas are more likely to avoid outdoor activities. This further complicates interpreting the results of the relationship between ambient concentrations and health to personal exposures.
Furthermore, variables that are expected to be confounders of personal exposures and health outcomes (such as an individual's smoking habits, variables $\widetilde{\bm C}_2$ in \cref{fig:outdoor_personal}) are not necessarily confounders of ambient \PM{} concentrations and outcomes (variables $\widetilde{\bm C}_1$ in \cref{fig:outdoor_personal}). Indication of confounding of the relationship between ambient concentrations and health outcomes by variables such as the median household value (see \cref{fig:post_inclusion}) implies that these variables are not confounders in the classic sense (since they are not driving ambient \PM{} concentrations) but are correlated with variables that are. Therefore, a variable's confounding strength for ambient concentrations is not directly interpretable as its confounding strength for personal exposures.

Second, this analysis has used log event rates as the outcome of interest in a linear regression setting, with all zip codes contributing equally to the estimation of the models irrespective of their size. Linear regression for the analysis of rate data has been used in various settings, for example in \cite{Joshua1990estimating, Mohamedshah1993truck, Chua2009pediatric, Wang2012} and \cite{LiuSmith2016gender}.
A Poisson regression model where the number of hospitalizations is the response variable and the Medicare population size within a zip code is the offset would be more in agreement with the literature on count outcomes (and within air pollution epidemiology specifically).
However, extending local confounding adjustment to Poisson regression is computationally complicated: (a) enforcing ER continuity for Poisson regression is not straightforward since $E_{\bm C}\{ E[Y | X = x, \bm C] \}$ is not easily acquired (see \cref{subsec:prior_continuity}), (b) posterior sampling of the experiment configuration and regression coefficient involves marginal densities of regression models, and (c) no conjugate prior distribution exists for regression coefficients in Poisson regression.

\bibliographystyle{plainnat}
\bibliography{Causal_ER,Air-pollution-review}


\appendix

\numberwithin{figure}{section}
\numberwithin{table}{section}
\addcontentsline{toc}{section}{Appendices}
\renewcommand{\thesection}{\Alph{section}}
\numberwithin{equation}{section}
\setcounter{figure}{0}    
\setcounter{table}{0}

\section{Data details}
\label{sec:data_details}
We constructed counts corresponding to the cardiovascular-specific (CVD) number of hospitalizations for Medicare enrollees aged at least 65 years during 2013 for 
zip codes across the continental US. Hospitalization rates were based on the total number of personal years for Medicare enrollees for a zip code on a given year. CVD hospitalizations were considered on the basis of primary diagnosis according to International Classification of Diseases, Ninth Revision (ICD-9) codes (ICD-9 390 to 459). The analysis was restricted to the continental US leading to 34,897 zip codes with hospitalization information.

Population demographic information was acquired using the 2000 Census with information on over 400 variables, although a lot of them are highly correlated. We further used linearly extrapolated Census variables for 2013. Census information is provided at a ZCTA level, and we use a crosswalk to map ZCTA to zip code. Weather information including temperature, relative humidity and dew point is acquired from the NOAA-ASOS (National Oceanic and Atmospheric Administration-Automated Surface Observing System) website, and is linked to zip codes within 150 kilometers.

Lastly, zip code PM$_{2.5}$ exposure is assigned using the US EPA monitoring sites. By EPA recommendations, monitoring sites with less than 67\% of scheduled measurements observed are excluded. For every monitor, the average of the 2011-2012 average annual value of PM$_{2.5}$ is calculated, and the monitor is linked to \textit{all} zip codes with centroids within 6 miles. Then, the zip code exposure is set equal to the average over all linked monitors.
Since monitoring sites are preferentially located near populated areas or points of interest, many zip codes in remote areas are not linked to any monitor and are therefore dropped from the final data set.

\begin{figure}
	\centering
	\begin{subfigure}[b]{0.45\textwidth}
		\includegraphics[width=\textwidth]{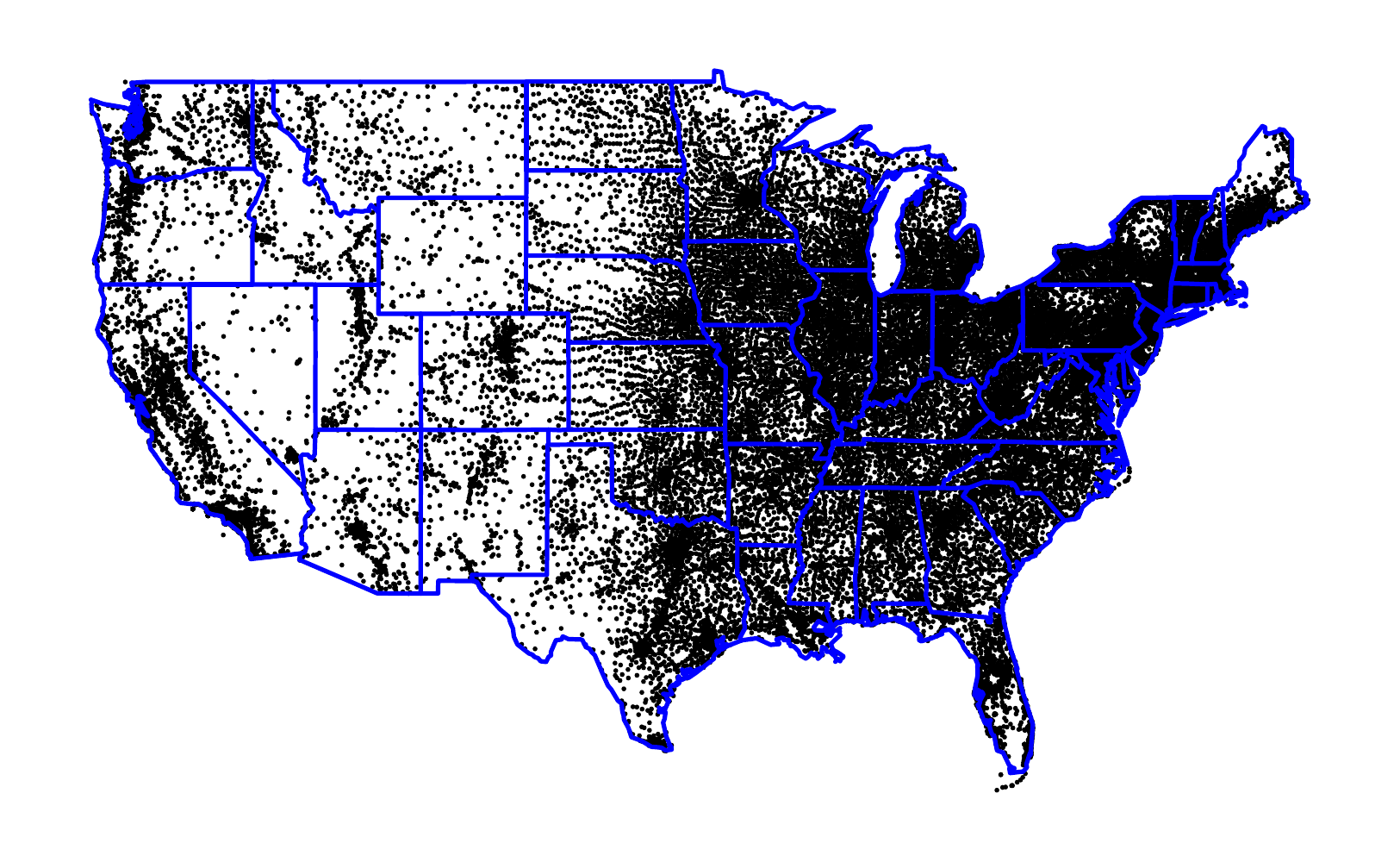}
		\caption{All zip codes in Medicare. \vspace{\baselineskip}}
	\end{subfigure} ~
	\begin{subfigure}[b]{0.45\textwidth}
		\includegraphics[width=\textwidth]{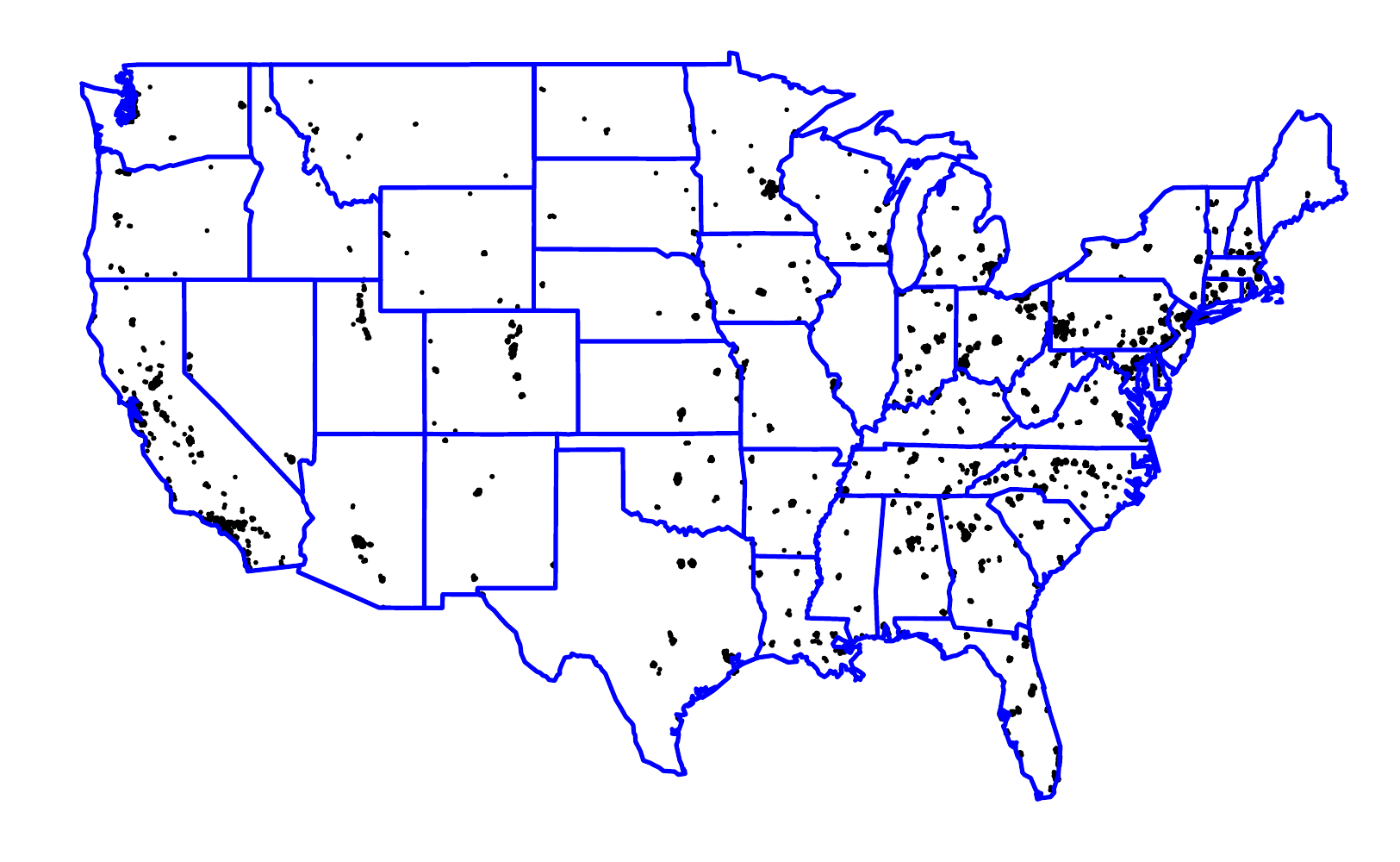} \caption{Zip codes with PM monitor within 6 miles. Linkage not unique.}
	\end{subfigure} ~
	\begin{subfigure}[b]{0.45\textwidth}
		\includegraphics[width=\textwidth]{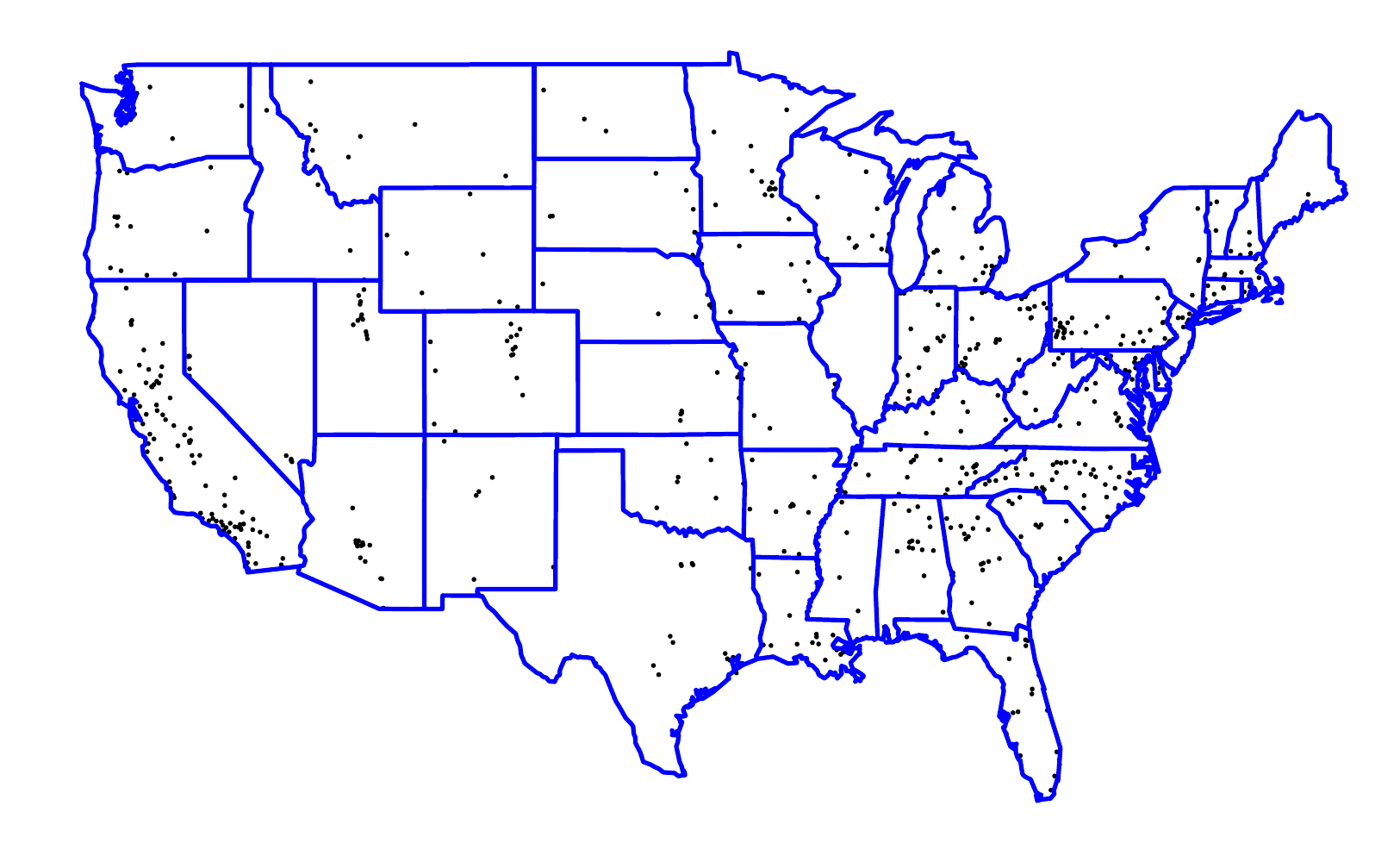} \caption{Zip codes with PM monitor within 6 miles. Unique linkage.}
	\end{subfigure} ~
	\begin{subfigure}[b]{0.45\textwidth}
		\includegraphics[width=\textwidth]{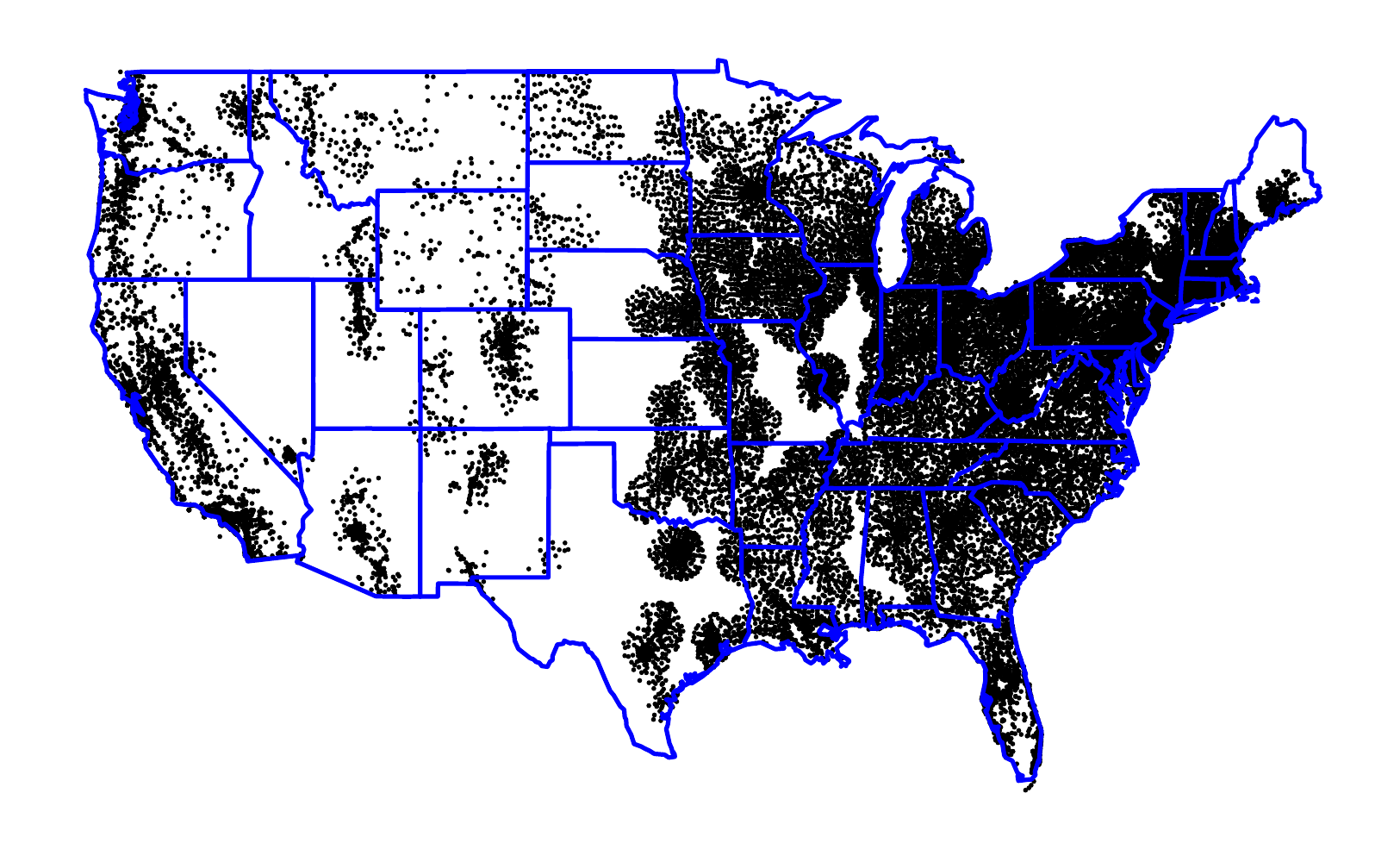} \caption{Zip codes with PM monitor within 60 miles. Linkage not unique.}
	\end{subfigure}
	\caption{(a) All zip codes with available Medicare information. (b) Zip codes with available exposure information after performing linkage within 6 miles and monitors are allowed to be linked to more than one zip code. (c) Zip codes with available exposure information after performing linkage within 6 miles where each monitor is only linked to up to one zip code. (d) Zip codes with available exposure information after linkage with monitors within 60 miles and every monitor can be linked to more than one zip code.}
	\label{fig:app_linkage}
\end{figure}

\cref{fig:app_linkage} shows maps of zip code centroids before linkage to EPA monitoring sites, as well as maintained zip code centroids after 3 different linkage procedures corresponding to different specifications of the linkage distance, as well as whether a monitor can be linked to more than one zip code. We visualize how linkage can affect the final data set:
\begin{itemize}
	\item \textbf{Distance:} As the distance of allowed linked zip codes and monitors increases, we expect that more zip codes will be linked to at least one monitor. However, PM$_{2.5}$ values in areas where monitors are located at long distances might suffer more from measurement error.
	\item \textbf{Number of links:} Allowing a monitor to be linked to multiple zip codes increases the number of zip codes with PM$_{2.5}$ information. However, this can lead to adjacent zip codes with very similar or identical PM$_{2.5}$ measurements.
\end{itemize}

Linkage (b) in \cref{fig:app_linkage} corresponds to the linkage procedure we used to create our data set. Table \ref{app_tab:Table1} includes the descriptive statistics and data source for the zip code covariates included in our analysis.

{\singlespacing
	\scriptsize
	\renewcommand*{\arraystretch}{1.4}
	\begin{longtable}{L{2cm}|L{2.3cm}|L{5.2cm}||c|c}
		\caption{Available demographic and weather information with mean (SD) in the sample and across all zip codes in the continental U.S.} \\
		\label{app_tab:Table1}
		Source & Name & Description & Sample & Full population$^*$ \\
		& & & 5,362 zip codes & 34,897 zip codes \\
		\hline
		2000 Census & \% White & Percentage of White Population & 0.71 (0.25) & 0.84  (0.2)  \\
		& \% Hisp & Percentage of Hispanic Population & 0.12 (0.18) & 0.07 (0.14)\\ 
		& \% HS &
		Percentage of population that attended high school & 0.27 (0.1) & 0.34  (0.11) \\ 
		& \% Poor & Percentage of impoverished population & 0.14 (0.11) & 0.13 (0.09) \\ 
		& \% Female & Percentage of female population & 0.51 (0.04) & 0.5 (0.04) \\ 
		&  \% Moved in 5 & Percentage of population that has lived in the area for less than 5 years & 0.50 (0.12) & 0.43 (0.12) \\
		& Avg Commute & Mean Travel Time to Work & 24.26 (6.02) & 26.11 (7.27) \\
		&  Population/SQM & Population per square mile (logarithm) & 7.52 (1.54) & 4.98 (2.22) \\
		&  Total Population & Total population (logarithm) & 9.7 (1.13) &
		8 (2.25) \\
		& Low Occupied & Indicator. ``=1'' if the percent of occupied population is at most 90\%. & 0.21 (0.41) & 0.44 (0.5) \\
		& High Occupied & Indicator. ``=1'' if the percent of occupied population is over 95\%. & 0.42 (0.49) & 0.24 (0.43) \\
		& Low Hispanic & Indicator. ``=1'' if the percent of Hispanic population is at most 0.02\% & 0.32 (0.47) & 0.54 (0.5) \\
		& High Hispanic & Indicator. ``=1'' if the percent of Hispanic population is over 20\% & 0.2 (0.4) & 0.1 (0.3) \\
		\hline
		Census Extrapolation
		& \% Below HS & Population percent with less than high school education (above age of 65) & 23.65 (15.11) & 23.4 (15.94) \\
		& \% Own Households & Percentage of occupied housing units in 2013 & 0.58 (0.2) & 0.72 (0.16) \\
		& Low Poverty & Indicator. ``=1'' if the percent of the population below the poverty line in 2013 is at most 5\% & 0.2 (0.4) & 0.28 (0.45) \\
		& High Poverty & Indicator. ``=1'' if the percent of the population below the poverty line in 2013 is over 15\%  & 0.24 (0.43) & 0.2 (0.4) \\
		\hline
		Census combination$^{**}$ &
		House Value & Median value of owner occupied housing (USD) (logarithm) & 12.65 (0.64) & 12.37 (0.62) \\
		& Household Income & Median household income (USD) (logarithm) & 11.41 (0.42) & 11.37 (0.36) \\ \hline
		BRFSS & BMI & Average BMI in 2013 & 27.67 (1.31) & 29.52 (4.81) \\
		& Smoking Rate & Ever smoke rate (2013) & 0.44 (0.06) &  0.46 (0.08) \\ \hline
		Weather &  Avg Temp & Average temperature (F) & 55.03 (7.14) & 54.34 (8.25) \\ 
		& Avg Dew Point & Average Dew Point (F) & 43.53 (6.96) & 43.41 (8.31) \\ 
		&  Avg Humidity & Average Relative Humidity (\%) & 69.93 (8.53) & 71.01 (8.21) \\
		\hline
		Medicare &  Avg Age & Average Medicare Age & 74.87 (1.62) & 74.51 (1.56) \\ 
		&  Female Rate & Percentage of Female Beneficiaries & 0.55 (0.06) & 0.53 (0.06) \\ 
		&  Dual Rate & Percentage of Dual Eligible Beneficiaries & 0.23 (0.15) & 0.18 (0.12) \\  \hline
	\end{longtable}
	\begin{flushleft}
		$^{*}$Missing values for covariates are excluded when calculating the mean and standard deviation in the full population. \\
		$^{**}$The 2000 Census is combined with the 2013 extrapolated values by taking each covariate's mean value across the two years.
	\end{flushleft}
}

\section{Additional results for investigation of local confounding}
\label{app_sec:local_confounding}

\subsection{Covariates' estimated coefficients in a model for the exposure at low and high exposure levels}

In \cref{sec:data_description}, we investigated the presence of local confounding by checking the p-values of each covariate in a model for the exposure, at low and high exposure levels separately. However, weak associations could have small p-values, if for example the sample size is large. Hence, investigations of local confounding could focus on the magnitude of estimated coefficients instead.

\begin{figure}[p]
	\centering
	\includegraphics[width=0.85\textwidth]{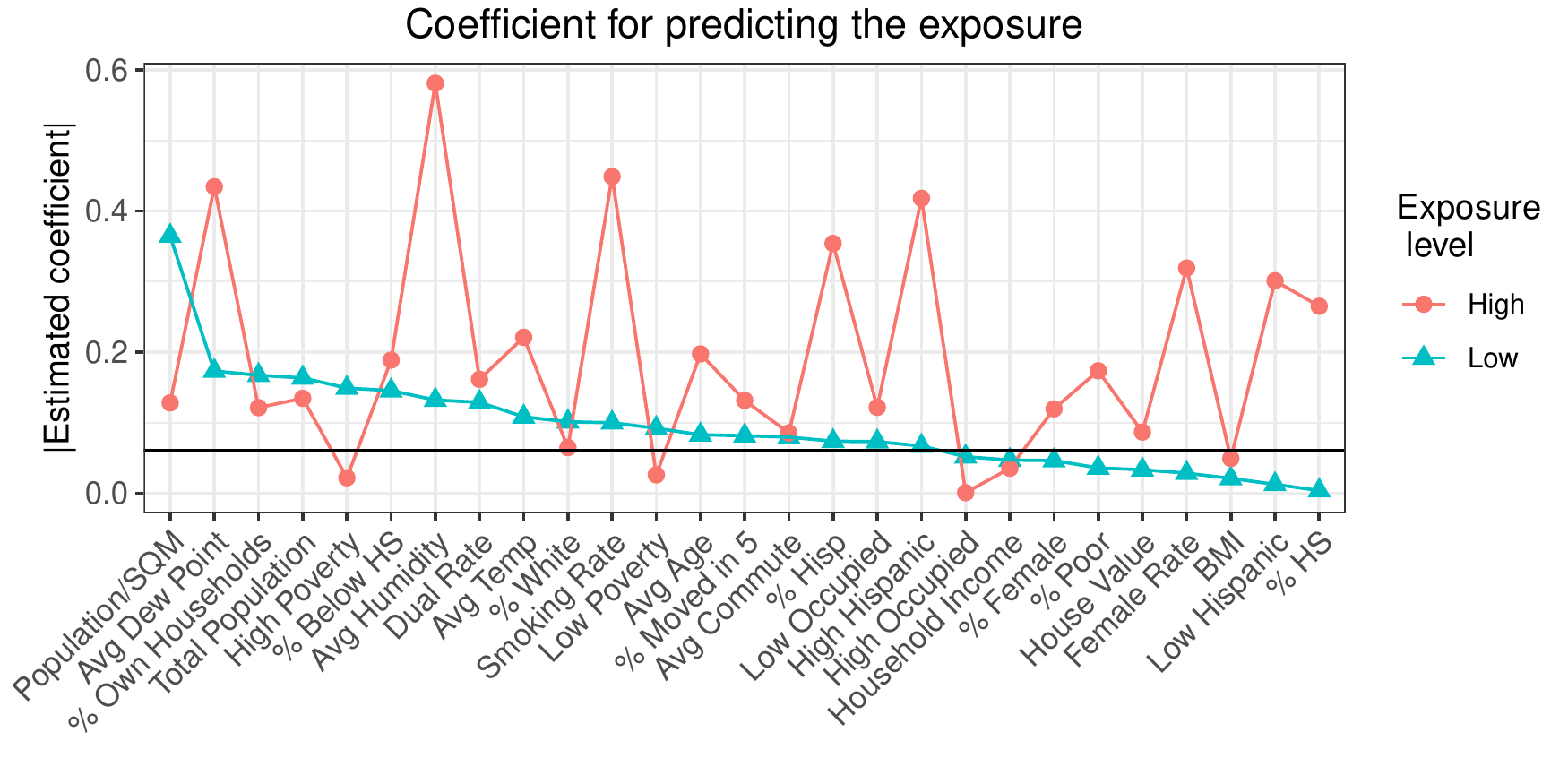}
	\caption{Absolute value of covariates' estimated coefficient in a regression of the exposure on each covariate separately and at the low and the high exposure levels.}
	\label{app_fig:app_coefs}
	\vspace{30pt}
	\includegraphics[width=0.85\textwidth]{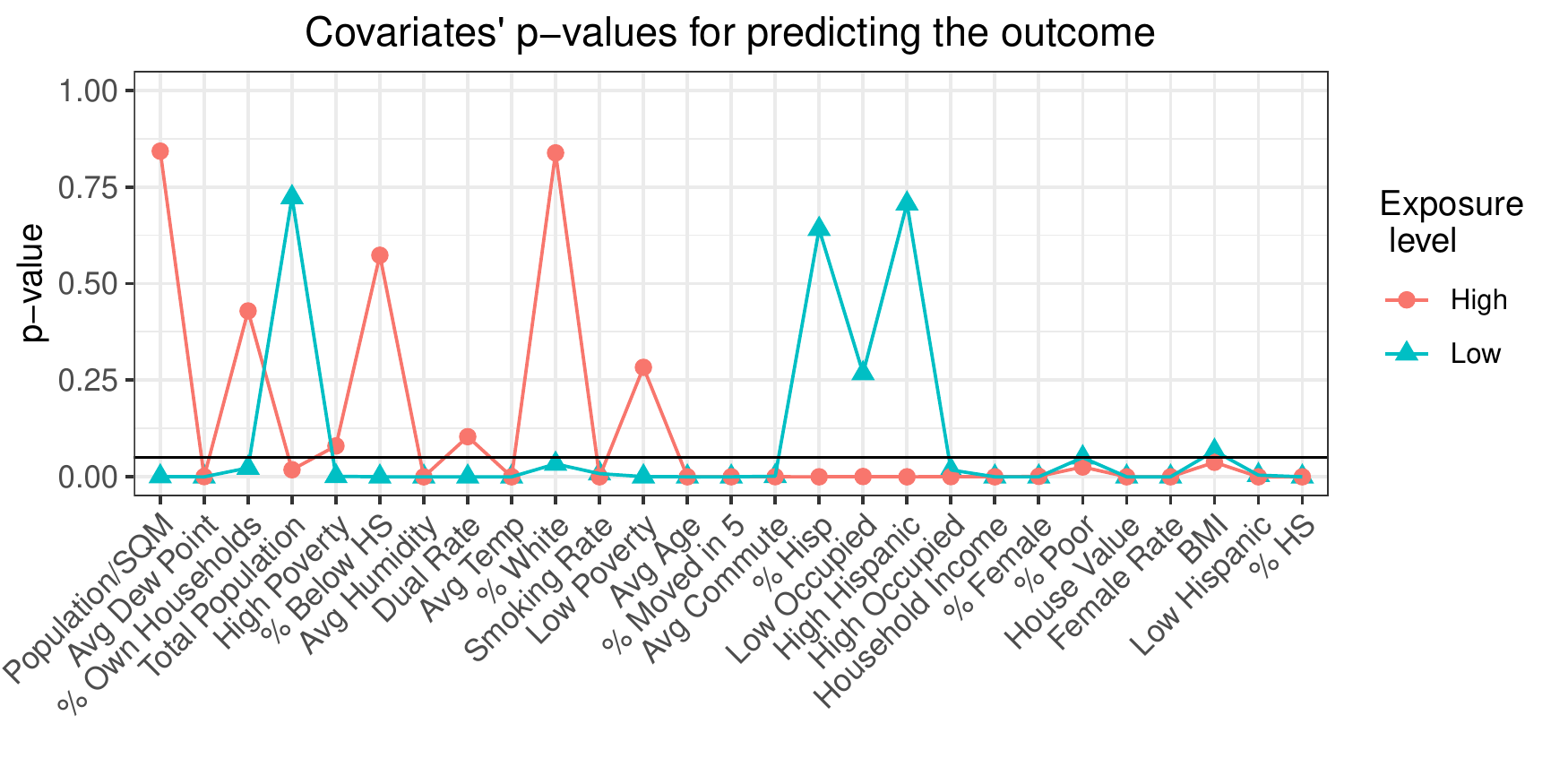}
	\caption{Covariates' $p$-values from regressing the outcome on each covariate separately. A separate regression is fit at the low and the high exposure levels.}
	\label{app_fig:app_out_pvalues}
\end{figure}

In \cref{app_fig:app_coefs}, we illustrate the absolute value of the estimated regression coefficients for the covariates in the regression model for predicting the exposure. In order for estimated coefficients to be comparable, covariates are standardized within each exposure level. The horizontal line corresponds to estimated coefficient equal to 0.06, a value between the coefficient for \texttt{High Hispanic} and \texttt{High Occupied}, covariates with p-value below and above 0.05 respectively. Comparing the estimated coefficients to the p-values in \cref{fig:app_pvalues} we see that there is a clear correspondence between small p-values and magnitude of estimated coefficient. That is expected when the two exposure levels have similar range of exposure values and similar number of observations.

This comparison indicates that when exposure levels are balanced, investigations of local confounding could focus on p-values or magnitude of estimated coefficients and lead to equivalent conclusions.

\subsection{Covariates' strength for predicting the outcome at the low and high exposure levels}

In \cref{sec:data_description}, we explored the possibility that different covariates are predictive of the exposure within different exposure levels. Here, we perform a similar analysis to investigate whether different variables act as predictors of the outcome at the low and the high exposure levels separately.

Considering still the two sets of zip codes: zip codes at the low exposure levels ($<$8$\mu g/m^3$) and zip codes at the high exposure levels ($>$11.5$\mu g/m^3$), we fit regressions of the outcome on the covariates, for each covariate and each exposure level separately. \cref{app_fig:app_out_pvalues} shows the covariates' $p$-value in those regressions. Again, we see that different variables act as predictors of the outcome at the two exposure levels. For example, a zip code's median house value (in logarithm -- \texttt{House Value}) is an outcome predictor at both low and high exposure levels, whereas the percentage of the population with less than high school education (\texttt{\% Below HS}) was an outcome predictor only at the low levels.

\section{Additional simulation results}
\label{app_sec:sims}

\subsection{Simulations in the presence of \textit{local} confounding}
\label{app_subsec:presented_sims}

\cref{app_table:sims_covs} shows the correlation of the covariates with the exposure and the coefficients of the covariates in the outcome model for the data simulating scenario with local confounding: different confounders at different levels of the exposure.

\begin{figure}[H]
	\centering
	\includegraphics[width = 0.85\textwidth]{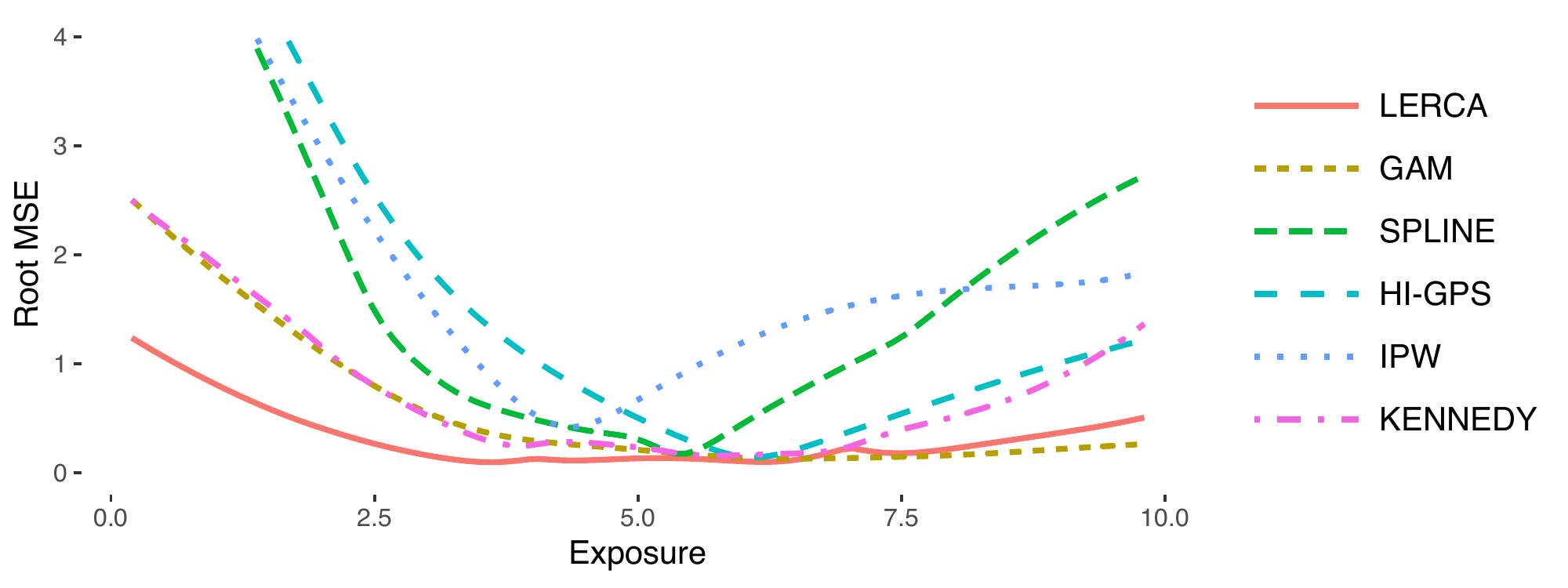}
	\caption{Mean Root MSE as a function of the exposure $x \in (0, 10)$.}
	\label{fig:sims_rootMSE}
\end{figure}

\cref{fig:sims_rootMSE} shows the the root MSE (rMSE) as a function of the exposure value $x\in(0, 10)$. LERCA has the lowest rMSE at the low exposure levels followed by GAM. Root MSE across most methods seems to be comparable for the middle exposure values, and GAM performs slightly better than LERCA at high levels.

\begin{table}[H]
	\caption{Correlation between the covariates and exposure, and outcome coefficients in each experiment, for scenarios with local confounding.}
	\label{app_table:sims_covs}
	\begin{tabular}{lr}
		\hspace{25pt} Covariate - Exposure \hspace{35pt} & \hspace{35pt} Covariate - Outcome
	\end{tabular}
	\centering
	\begin{tabular}{r||rrrr||rrrr}
		\hline
		& $x\in g_1$ & $x\in g_2$ & $x\in g_3$ & $x\in g_4$ & $x\in g_1$ & $x\in g_2$ & $x\in g_3$ & $x\in g_4$  \\ 
		\hline
		$C_1$    & 0.423 & 0.525 & 0.402 & 0      &
		0.641 & 0     & 0     & 0      \\ 
		$C_2$    & 0.524 & 0.572 & 0     & 0.503  &
		0.962 & 0.919 & 0.593 & 0.651  \\
		$C_3$    & 0.522 & 0     & 0.447 & 0      & 
		0.646 & 0.643 & 0.616 & 0.58  \\
		$C_4$    & 0     & 0.528 & 0     & 0      &
		0     & 0.633 & 0     & 0      \\ 
		$C_5$    & 0     & 0     & 0.533 & 0.539  &
		0     & 0     & 0.658 & 0      \\ 
		$C_6$    & 0     & 0     & 0     & 0.509  &
		0     & 0     & 0     & 0.52   \\ 
		$C_7$    & 0     & 0     & 0     & 0      &
		0     & 0     & 0     & 0      \\ 
		$C_8$    & 0     & 0     & 0     & 0      &
		0     & 0     & 0     & 0     \\ 
		\hline
	\end{tabular}
\end{table}

\vspace{20pt}

\subsection{Simulations with reversed confounding structure}
\label{app_subsec:sims_local_reverse}

In \cref{fig:sims_rootMSE}, we see that GAM has slightly lower root MSE than LERCA at high exposure levels. A natural question that arises is whether the relative performance of GAM and LERCA is (1) a feature of the specific confounding structure used to generate the data in those simulations, or (2) specific to the high exposure range. For that reason, we generated data over the same exposure range $(0, 10)$, the same experiment configuration $(0, 2, 4, 7, 10)$, but reversed confounding structure. Specifically, the confounder set and confounding strength of experiment 1 in the simulations presented here are those of experiment 4 in the original simulations, those of experiment 2 here are those of experiment 3 in the original simulations, and so forth.

\begin{figure}[H]
	\centering
	\includegraphics[width=\textwidth]{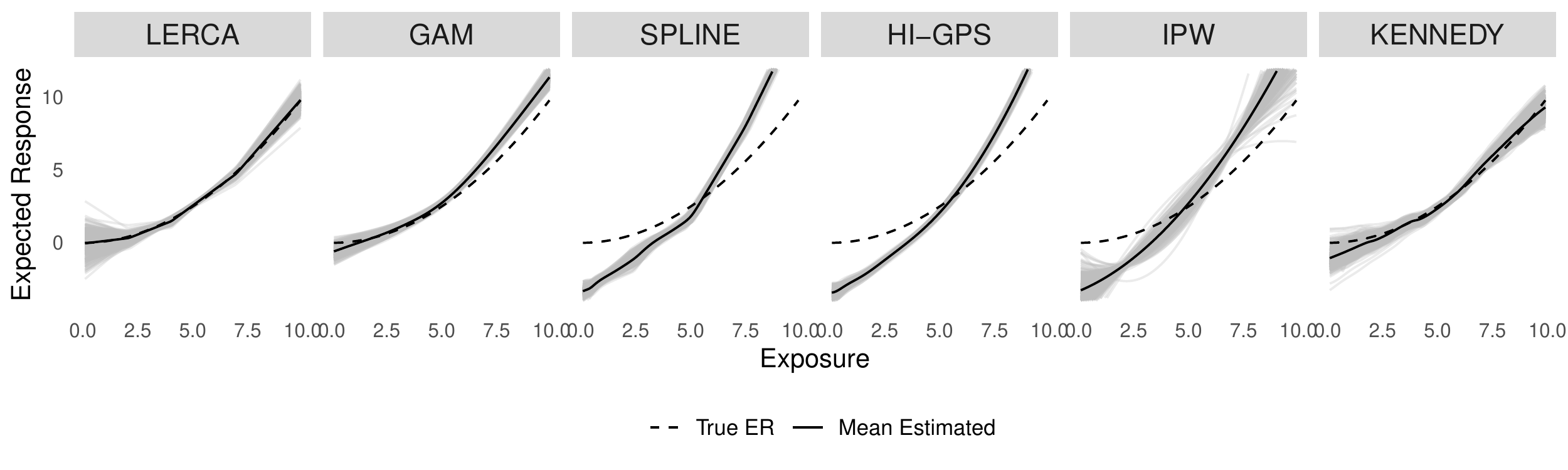}
	\vspace{20pt}
	\includegraphics[width=0.8\textwidth]{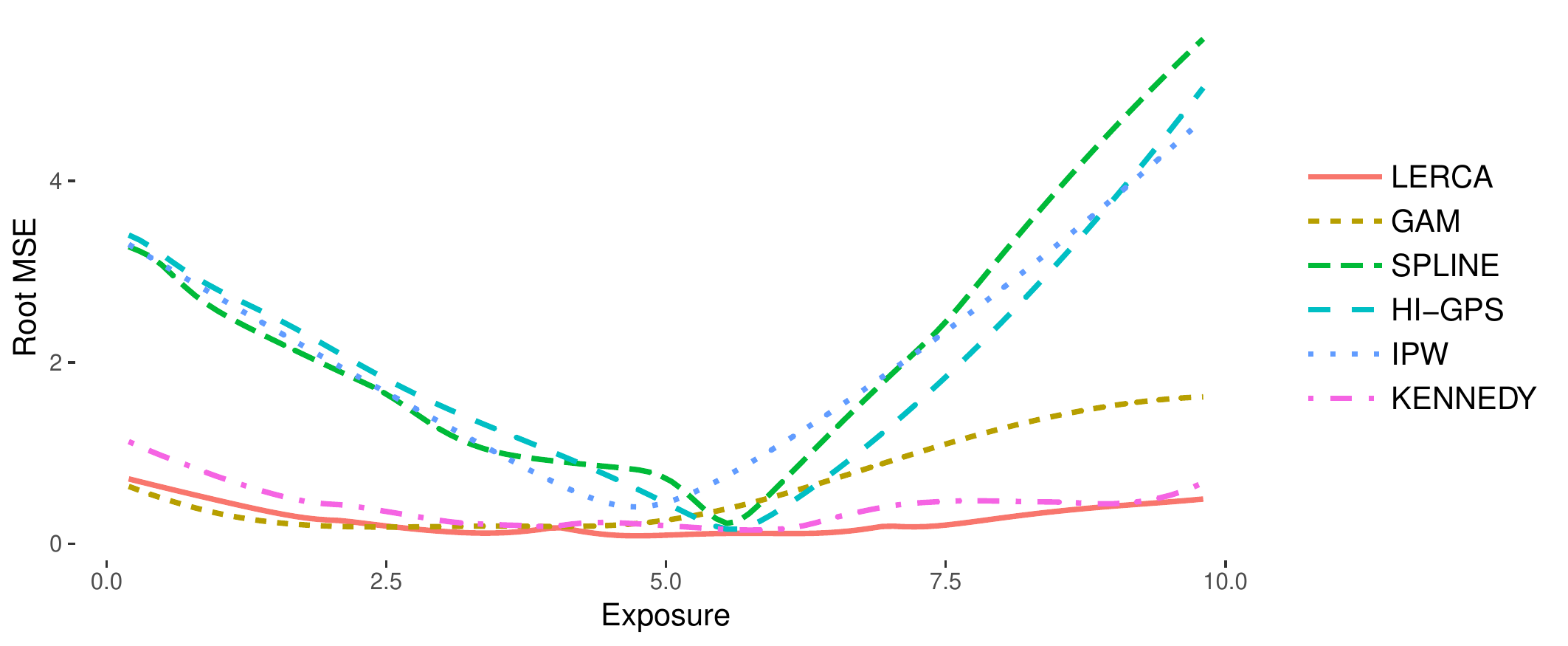}
	\caption{Simulation results over 400 data sets with reversed confounding structure.}
	\label{app_fig:sims_reverse}
\end{figure}

\cref{app_fig:sims_reverse} show the estimated ER curves and root MSE for all methods. For computational simplicity, we only considered LERCA with $K = 3$. In these results we see that the estimated ER using LERCA follows the true ER curve very closely, whereas the remaining of the methods have deviations from the true ER curve at low or high exposure levels. When considering the root MSE of the methods, we see that GAM slightly outperforms LERCA at low exposure levels, but LERCA does much better at high exposure levels. Combining these results to those in \cref{fig:sims_rootMSE}, we see that (1) the performance of GAM relative to LERCA highly depends to the confounding structure, and (2) LERCA achieves the lowest or close to the lowest root MSE across exposure levels and under either specification of confounding structure.

\subsection{Simulations in the presence of \textit{global} confounding}
\label{app_subsec:sims_global}

\begin{table}[!b]
	\centering
	\caption{Correlation between the covariates and exposure, and outcome coefficients in each experiment, for the scenario with global confounding.}
	\label{app_table:sims_covs2}
	\begin{tabular}{r|rrrrrrrr}
		& $C_1$ & $C_2$ & $C_3$ & $C_4$ & $C_5$ & $C_6$ & $C_7$ & $C_8$ \\
		\hline
		Exposure & 0.423 & 0.524 & 0.522 & 0 & 0 & 0 & 0 & 0 \\
		Outcome & 0 & 0.812 & 0.93 & 0.82 & 0 & 0 & 0 & 0 \\
		\hline
	\end{tabular}
\end{table}

\begin{figure}[!t]
	\centering
	\vspace{20pt}
	\includegraphics[width = \textwidth]{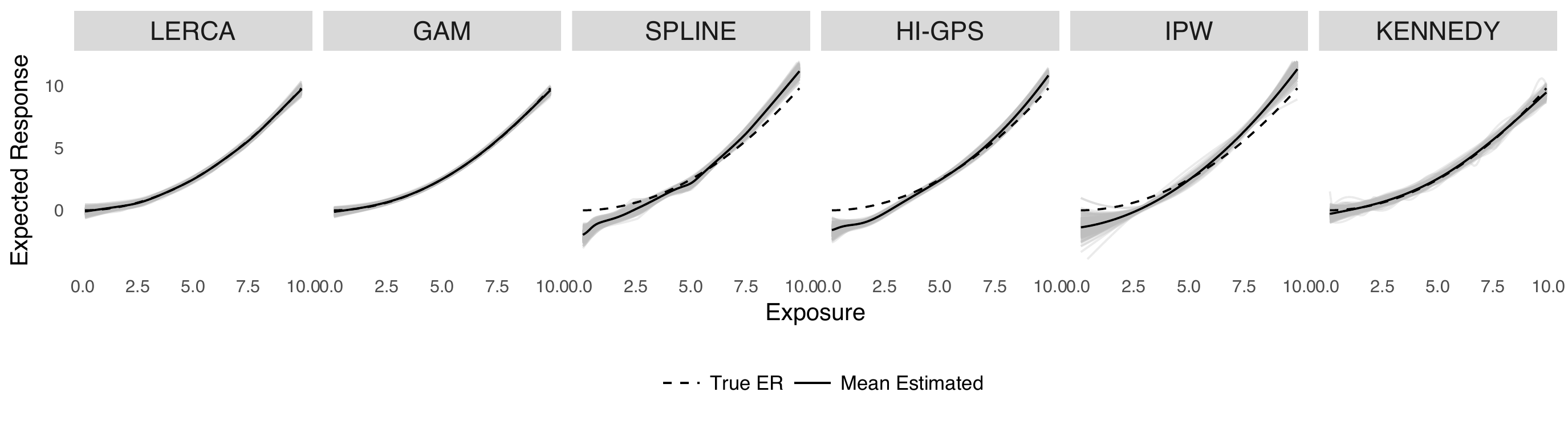}
	\caption{Simulation results in the presence of global confounding. Grey lines correspond to estimated ER for each simulated data set, solid lines correspond to the mean ER over all simulated data sets, and the dashed line corresponds to the true ER.}
	\label{app_fig:sims_global}
	\centering
	\includegraphics[width = 0.85\textwidth]{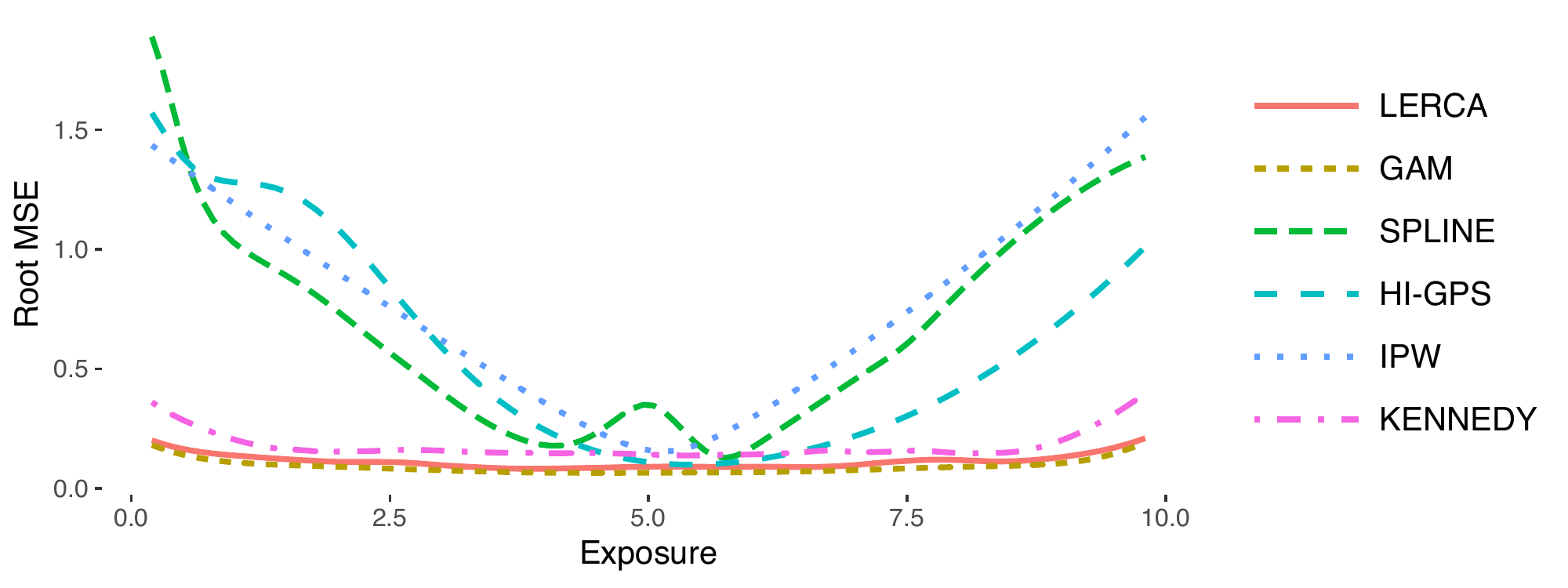}
	\caption{Root MSE of all methods in the presence of global confounding as a function of the exposure $x\in (0, 10)$.}
	\label{app_fig:sims_global_mse}
\end{figure}

Briefly, data are generated with covariates $C_1, C_2, C_3$ as predictors of exposure and $C_2, C_3, C_4$ as predictors of the outcome and the adjusted R-squared of the true exposure and outcome models was 0.73 and 0.94 accordingly.
\cref{app_table:sims_covs2} shows the correlation of covariates with the exposure and the outcome model coefficients in the data simulating scenario with global confounding (same confounders with constant confounding strength across exposure levels) and true quadratic ER.
\cref{app_fig:sims_global} shows the estimated ER for each data set and the average estimated ER based on LERCA and alternative methods. In \cref{app_fig:sims_global_mse}, the root MSE for all methods is plotted as a function of the exposure $x \in (0, 10)$.

\section{Prior specifications for regression parameters and experiment configuration}
\label{app_sec:priors}

\subsection{Regression coefficients and residual variance}
\label{app_subsec:prior_coef}

Prior independences of all parameters are expressed in the following representation
\begin{equation}
\label{app_eq:prior}
\begin{aligned}
& p(\tend{\alpha}^X, \tend{\alpha}^Y, \tend{\delta}^X, \beta_k, \tend{\delta}^Y, \tend\sigma^2_{X}, \tend\sigma^2_{Y}) \\
&\ = p(\delta_{10}^Y) \ \prod_{k = 1}^{K + 1}\left\{
\left[ \prod_{j = 1}^p p(\alpha_{kj}^X, \alpha_{kj}^Y)\ 
p(\delta_{kj}^X | \alpha_{kj}^X) \
p(\delta_{kj}^Y| \alpha_{kj}^Y) \right]
p(\delta_{k0}^X) \ p(\beta_k) \ 
p(\sigma^2_{k, X}) \ 
p(\sigma^2_{k, Y})
\right\} \\
& \hspace{35pt} \times \prod_{k = 2}^K I\left(\delta_{k0}^Y = \delta_{(k - 1)0}^Y + \beta_{k - 1}(s_k - s_{k-1}\right).
\end{aligned}
\end{equation}
We assume non-informative normal priors on $\beta_k$, $k = 1, 2, \dots, K + 1$, and $\delta_{10}^Y$. The prior distribution on the regression coefficients for the covariates is a mixture of non-informative normal distribution and point-mass at 0. Non-informative inverse gamma prior distributions are assumed on $\sigma_{k,X}^2, \sigma_{k,Y}^2$. Specifically
\begin{itemize}
	\item $\beta_k \sim N(\mu_0, \sigma^2_0)$, $\delta_{10} \sim N(\mu_0, \sigma^2_0)$.
	\item $\delta_{kj}^X | \alpha_{kj}^X \sim \alpha_{kj}^X \ 
	N(\mu_0, \sigma^2_0) + (1 - \alpha_{kj}^X) \ 
	\mathbb{1}_0(\delta_{kj}^X),$ where
	$ \mathbb{1}_0(\delta_{kj}^X)$ is a point-mass distribution at 0. 
	Similarly for $\delta_{kj}^Y | \alpha_{kj}^Y$.
	\item $\sigma^2_{k, X} \sim IG(a_0, b_0)$, and similarly for $\sigma^2_{k, Y}$.
\end{itemize}
The hyper-parameters $\mu_0, \sigma^2_0, a_0, b_0$ can be chosen differently for different variables.

\subsection{Experiment configuration}
\label{app_subsec:prior_s}

The prior on the points $\mb s = (s_1,s_2, \dots, s_K)$ defining the experiment configuration is set as the even ordered statistics of $(2K + 1)$ samples from a uniform distribution over the observed exposure range. Compared to a uniform prior distribution on $\mb s$, this choice of a prior discourages the existence of points $s_i, s_j$ in the experiment configuration that are very close to each other.

Let $K$ and the exposure range $(s_0, s_{K + 1})$ be fixed. Let $Z_i \sim U(s_0, s_{K + 1})$, $i = 1, 2, \dots, 2K + 1$ and denote the even ordered statistics as $W_j = Z_{(2j)}, j = 1, 2, \dots, K$. Then,
\begin{align*}
f_{W_1, W_2, \dots, W_K}&(w_1, w_2, \dots, w_K) = \\ 
& f_{W_1}(w_1) f_{W2 | W1}(w_2 | w_1) \dots f_{W_K | W_1, W_2, \dots, W_{K - 1}}(w_k | w_1, w_2, \dots, w_{k = 1})
\end{align*}
Since $W_1$ is th 2$^{nd}$ order statistic of $2K + 1$ samples from $U(s_0, s_{K + 1})$, we know that
\begin{align*}
f_{W_1}(w_1) = & \frac{(2K + 1)!}{(2K - 1)!} \frac1{s_{K + 1} - s_0}
\frac{w_1 - s_0}{s_{K + 1} - s_0}
\left(1 - \frac{w_1 - s_0}{s_{K + 1} - s_0} \right)^{2K - 1} \\
= & \frac{(2K + 1)!}{(2K - 1)!} (s_{K + 1} - s_0)^{-(2K + 1)}
(w_1 - s_0)(s_{K + 1} - w_1)^{2K - 1}
\end{align*}
Given $W_1 = w_1$, $W_2$ acts like the second order statistic of $2K - 1$ uniform samples from a uniform distribution over $(w_1, s_{K + 1})$. Therefore, we similarly get that
\begin{align*}
f_{W_2 | W_1}(w_2 | w_1) = &
\frac{(2K - 1)!}{(2K - 3)!} (s_{K + 1} - w_1)^{-(2K - 1)}
(w_2 - w_1)(s_{K + 1} - w_2)^{2K - 3}.
\end{align*}
Iteratively, we have that
\begin{align*}
f_{W_1, W_2, \dots, W_K}&(w_1, w_2, \dots, w_K) = \\ 
& (2K + 1)!(s_{K + 1} - s_0)^{-(2K + 1)}(w_1 - s_0)(w_2 - w_1) \dots (w_K - w_{K - 1})(s_{K + 1} - w_K).
\end{align*}
Therefore, the prior distribution on $\mb s$ with minimum distance of consecutive points $s_k, s_{k + 1}$ being $d_k$ is defined as
\begin{align}
f_{\mb s}(s_1, s_2, \dots, s_K) \propto \prod_{k = 0}^K
(s_{k+1} - s_k) \mathbb{1}(s_{k+1} - s_k > d_k)
\label{app_eq:prior_experiments}
\end{align}

\section{Sampling from the posterior distribution}
\label{app_sec:MCMC}

The parameters included in the model are:
$\mb s$ (the exposure values in the experiment configuration),
$\tend{\alpha}^X$, $\tend{\alpha}^Y$ (the vectors of length $p$ including the covariates' inclusion indicators in the exposure and the outcome model for each experiment),
$\tend{\beta} = \{\beta_k\}_{k = 1}^{K + 1}$ (coefficients of exposure in the outcome model),
$\tend{\delta}^X, \tend{\delta}^Y$ (intercepts and coefficients of the covariates in the exposure and outcome model of each experiment), 
$\tend\sigma_X^2 = \{\sigma_{k, X}^2\}_{k = 1}^{K + 1}, \tend\sigma^2_Y = \{\sigma^2_{k,Y}\}_{k = 1}^{K + 1}$ (residual variance of the exposure and outcome within each experiment).

\subsection{Likelihood factorization}
\label{app_subsec:factorization}

We start by noting that the data likelihood (conditional on all parameters) factorizes to components for different experiments and the exposure and outcome models. If $\mb Y, \mb X$ denote the vectors of outcomes and exposures for all units in the sample, and $\mb Y^k, \mb X^k$ denote the vectors of outcomes and exposures in experiment $k$, then
\begin{align*}
P(\mb{Y}, \mb{X} | & \mb{s}, \tend\alpha^X, \tend\alpha^Y, \tend\delta^X, \tend\delta^Y, \tend\beta, \tend\sigma^2_X, \tend\sigma^2_Y, \mb C)
= \\
& \prod_{k = 1}^{K + 1} \prod_{i \in g_k}
p_k(Y_i |X_i, \tend\alpha_k^Y, \tend\delta_k^Y, \beta_k, \sigma^2_{k,Y}, \mb{C}_i) p_k(X_i | \tend\alpha_k^X, \tend\delta_k^X, \sigma^2_{k,X}, \mb{C}_i) = \\
& \prod_{k = 1}^{K + 1} \left[ 
p_k(\mb Y^k | \mb X^k, \tend\alpha_k^Y, \tend\delta_k^Y, \beta_k, \sigma^2_{k,Y}, \mb{C}^k) p_k(\mb X^k | \tend\alpha_k^X, \tend\delta_k^X, \sigma^2_{k,X}, \mb{C}^k) \right],
\numberthis
\label{app_eq:factor_conditional}
\end{align*}
where we denote $p_k(\cdot_1 | \cdot_2)$ as the density of $\cdot_1$ conditional on $\cdot_2$ in experiment $k$ and $\tend\delta_k^Y$ includes the intercept $\delta_{k0}^Y$.

Next, we note that if we consider the marginal likelihood integrating out 1) exposure model regression coefficients including the intercept, 2) outcome model covariates' regression coefficients, and 3) all variance terms, then the likelihood still factorizes in a similar manner. In fact\footnote{In the following, $\tend\delta^X$ includes the exposure model intercepts, but $\tend\delta^Y$ includes only the coefficients of the covariates.}:
\begin{align*}
P(&\mb Y, \mb X |\mb s, \tend\alpha^X, \tend\alpha^Y, \tend\beta, \delta_{10}^Y, \mb C)\\
= & \int P(\mb Y, \mb X |\mb s, \tend\alpha^X, \tend\alpha^Y, \tend\delta^X, \tend\delta^Y, \tend\beta, \delta_{10}^Y, \tend\sigma^2_X, \tend\sigma^2_Y, \mb C ) \times \\
& \hspace{40pt} p(\tend\delta^X, \tend\delta^Y, \tend\sigma^2_X, \tend\sigma^2_Y | \mb s, \tend\alpha^X, \tend\alpha^Y)\ 
\mathrm{d}(\tend\delta^X, \tend\delta^Y, \tend\sigma^2_X, \tend\sigma^2_Y) \\
= & \prod_{k = 1}^{K + 1} \int
p_k(\mb Y^k | \mb X^k, \mb s, \tend\alpha_k^Y, \tend\delta_k^Y, \tend\beta, \delta_{10}^Y, \sigma^2_{k,Y}, \mb{C}^k) p_k(\mb X^k | \tend\alpha_k^X, \tend\delta_k^X, \sigma^2_{k,X}, \mb{C}^k) \times \\
& \hspace{40pt}
p(\tend\delta_k^X, \tend\delta_k^Y, \sigma^2_{k,X}, \sigma^2_{k,Y} | \mb s, \tend\alpha_k^X, \tend\alpha_k^Y) \ \ 
\mathrm{d}(\tend\delta_k^X, \tend\delta_k^Y, \sigma^2_{k,X}, \sigma^2_{k,Y}) \\
= &
\prod_{k = 1}^{K + 1} \int p_k (\mb Y^k | \mb X^k, \mb s, \tend\alpha_k^Y, \tend\delta_k^Y, \tend\beta, \delta_{10}^Y, \sigma^2_{k,Y}, \mb C^k)
p(\tend\delta_k^Y,  \sigma^2_{k,Y} | \mb s, \tend\alpha_k^Y)\ \mathrm{d}(\tend\delta_k^Y, \sigma^2_{k,Y}) \\
& \hspace{0.7cm}
\int p_k (\mb X^k | \tend\alpha_k^X, \tend\delta_k^X, \sigma^2_{k,X}) p(\tend\delta_k^X, \sigma^2_{k,X} | \mb s, \tend\alpha_k^X)\ \mathrm{d}(\tend\delta_k^X, \sigma^2_{k,X}) \\
=& 
\prod_{k=1}^{K + 1}
p_k (\mb Y^k | \mb X^k, \mb s, \tend\alpha_k^Y, \delta_{k0}^Y, \beta_k, \mb C^k)
p_k (\mb X^k | \tend\alpha_k^X, \mb C^k)
\numberthis
\label{app_eq:marginal_factorization}
\end{align*}
where the second equation holds from the factorization of the likelihood (when $\tend \delta_k^Y$ does not include the intercepts we need to condition on $\delta_{10}^Y$ and $\tend{\beta}$) and the assumed prior independences.


\subsection{Sampling all model parameters using MCMC}
\label{app_subsec:MCMC}

\subsubsection{Sampling the regression coefficients and residual variance terms}
\label{app_subsec:sample_coefs}

The factorization of the full data likelihood over experiments and exposure/outcome models and the choice of the prior distributions lead to full conditional posterior distributions of coefficients $\delta_{k0}^X, \delta_{kj}^X, \delta_{kj}^Y$, and variance terms $\sigma^2_{k,X}, \sigma^2_{k, Y}$ of known forms. The variance terms and exposure model intercepts have inverse Gamma and normal full conditional posterior distributions accordingly, whereas the distributions of $\delta_{kj}^X, \delta_{kj}^Y$ are either point mass at 0 or normal, based on whether the corresponding $\alpha$ is 0 or 1.

We update coefficients $\delta_{kj}^X$ for which $\alpha_{kj}^X = 0$ separately from the ones with $\alpha_{kj}^X = 1$.
Parameters $\delta_{kj}^X$ for which $\alpha_{kj}^X = 0$ are set to 0.
Let $j_1, j_2, \dots, j_{N_x}$ be the indices such that $\alpha_{kj_l} = 1$, $l = 1, 2, \dots, N_x$. Then,
\begin{align*}
& (\delta_{k0}^X, \delta_{kj_1}^X, \delta_{kj_2}^X, \dots, \delta_{kj_{N_x}}^X) ^T | \text{Data}, \mybullet \sim
MVN_{N_x + 1} (\mu_X, \Sigma_X), \\
\text{ where } &
\Sigma_X = \left( \frac1{\sigma^2_{k, X}} \tilde{\mb V}^T \tilde{\mb V}
+ \frac1{\sigma^2_0} I_{N_x + 1} \right)^{-1} \ \text{and} \ 
\mu_X = \Sigma_X \left( \frac{1}{\sigma^2_{k,X}}
\tilde{\mb V}^T \mb X^k + \frac1{\sigma^2_0}\tilde\mu_0 \right)
\end{align*}
where $\tilde{\mb V} = (\mb 1, \mb C^k_{j_1}, \mb C^k_{j_2}, \dots, \mb C^k_{j_{N_x}})$ is the design matrix of data in experiment $k$ based on the included covariates, and $\tilde \mu_0$ is a vector of length $N_x + 1$ of repeated values $\mu_0$. (Update of the coefficients $\delta_{kj}^Y$ is performed conditional on $\delta_{k0}^Y, \beta_k$ and is similar to the updates of the coefficients in the exposure model and therefore omitted.)

The full conditional distribution of the variance term $\sigma^2_{k,X}$ is also of known form
\begin{align*}
& \sigma^2_{k,X} |\text{Data}, \mybullet \sim IG(a_X, b_X), \\
\text{ where } & a_X = a_0 + \frac{n_k}2, \ \ 
b_X = b_0 + \frac12 (\mb X^K - \mb V \tend\delta_k^X)^T
(\mb X^k - \mb V\tend\delta_k^X),
\end{align*}
where $n_k$ is the number of observations in experiment $k$, and $\mb V = (\mb 1, \mb C^k)$.
(The full conditional posterior distribution of $\sigma^2_{k,Y}$ is very similar and is therefore omitted.)

It is worth noting that centering the covariates $C_j$ allows the outcome model intercepts $\delta_{k0}$ to depend solely on $\delta_{10}$, $\beta_k$ and $\mb s$, and not on $\delta_{kj}^Y$. This simplifies the form of the full conditional distribution for many coefficients. Since $\delta_{k0}^Y$, $k \geq 2$ is a deterministic function of $\delta_{10}$, $\beta_1, \beta_2, \dots, \beta_{k-1}$, and the points $s_0, s_1, \dots, s_k$, the full conditional posterior distribution of $\delta_{10}$ depends on data across all experiments, and that of $\beta_k$ on data from experiment $k$ and onwards. Since the data likelihood in all experiments is normal and we have assumed normal prior distributions, the full conditional posterior distributions are also normal.
After each update, intercepts $\delta_{k0}^Y, k \geq 2$ need to be updated from \cref{eq:intercept_prior} to ensure ER continuity.

The parameter $\delta_{10}^Y$ is included in the mean structure of the outcome model for all experiments. Its full conditional posterior distribution is $\delta_{10}^Y | \text{Data}, \mybullet \sim N(\mu, \sigma^2)$ where
$$
\sigma^2 = \left(\frac1{\sigma^2_0} + \sum_{k = 1}^{K+1} \frac{n_k}{\sigma^2_{k,Y}} \right)^{-1}
$$
and
$$
\mu = \sigma^2 \left[
\frac{\mu_0}{\sigma^2_0} +
\sum_{k = 1}^{K + 1} \frac1{\sigma^2_{k, Y}} \sum_{i\in g_k} \left(
y_i - \sum_{l = 1}^{k - 1}\beta_l(s_l - s_{l - 1}) - \beta_k(x_i - s_{k - 1}) - \sum_{j = 1}^p \delta_{kj}^Y C_{ij}
\right) \right],
$$
where $\sum_a^b = 0$ if $b < a$.
Similarly, the full conditional posterior distribution of $\beta_k$ uses data from experiments $k, k + 1, \dots, K + 1$, and is $\beta_k | \text{Data}, \mybullet \sim N(\mu, \sigma^2)$ where
$$
\sigma^2 = \left( \frac1{\sigma_0^2} + \frac1{\sigma^2_{k,Y}} \sum_{i \in g_k} (x_i - s_{k - 1})^2 + (s_k - s_{k - 1})^2 \sum_{l = k+1}^{K + 1}\frac{n_l}{\sigma^2_{l,Y}} \right)^{-1}
$$
and
{\small
	\begin{align*}
	\mu = & \sigma^2 \left( \frac{\mu_0}{\sigma^2_0} + \frac1{\sigma^2_{k,Y}} \sum_{i \in g_k}(x_i - s_{k -1}) \big(y_i - \delta_{k0}^Y - \textstyle{\sum_j} \delta_{kj}^Y C_{ij}\big) + \right. \\
	&  \hspace{20pt} \left.
	\sum_{l = 1}^{K + 1} \frac{1}{\sigma^2_{l,Y}} \sum_{i \in g_l} (s_k - s_{k - 1})(y_i - \delta_{k0}^Y - \textstyle\sum_{e = k + 1}^{l - 1}\beta_e(s_e - s_{e - 1}) - \beta_l(x_i - s_{l - 1}) - \textstyle{\sum_j} \delta_{lj}^Y C_{ij})
	\right).
	\end{align*}
}

\subsubsection{Sampling the experiment configuration and inclusion indicators}
\label{app_subsec:sample_cuts}

\begin{figure}[!t]
	\centering
	\begin{subfigure}[b]{0.23\textwidth}
		\includegraphics[width=\textwidth]{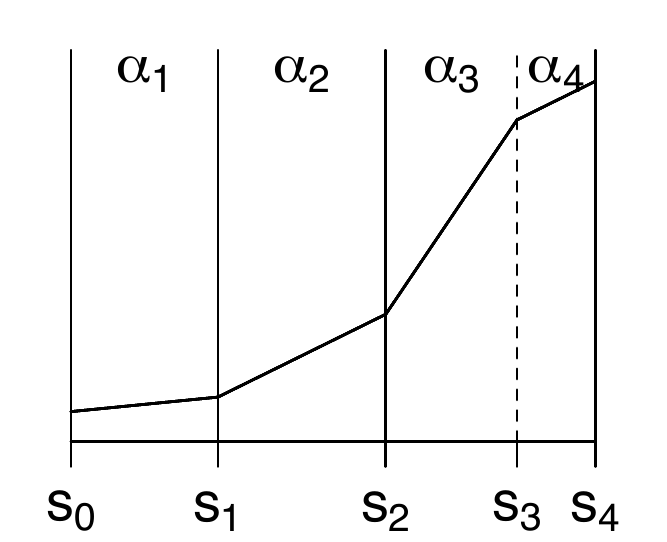}
		\caption{Current state}
	\end{subfigure}
	~
	\begin{subfigure}[b]{0.23\textwidth}
		\includegraphics[width=\textwidth]{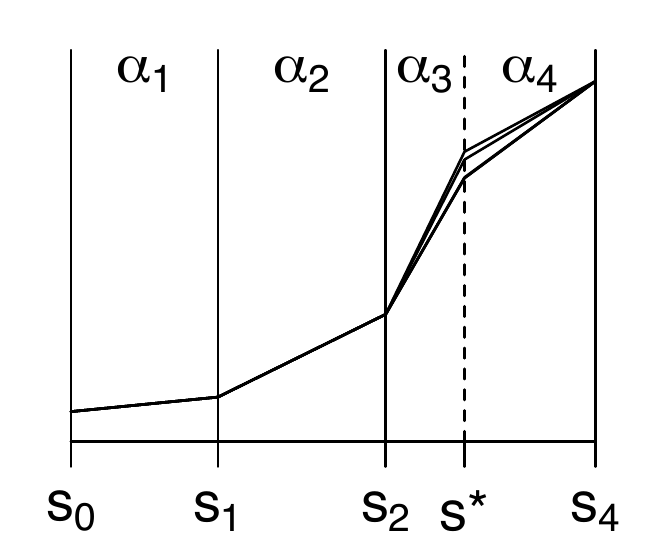}
		\caption{Separate}
		\label{subfig:MCMC_separate}
	\end{subfigure}
	~
	\begin{subfigure}[b]{0.23\textwidth}
		\includegraphics[width=\textwidth]{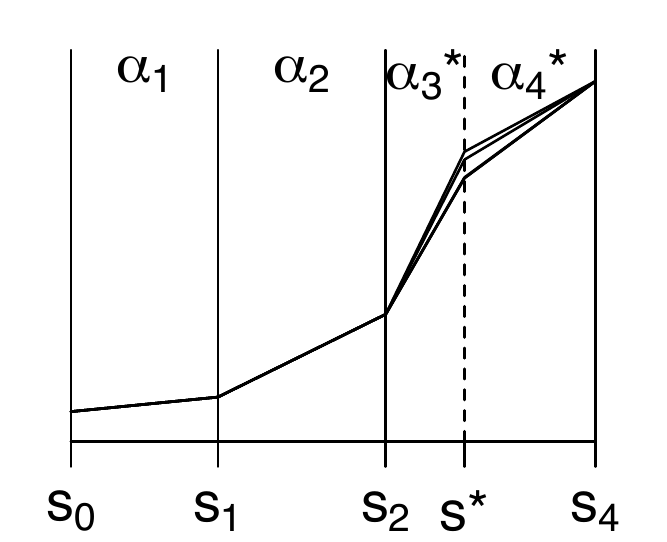}
		\caption{Jump within}
		\label{subfig:MCMC_jump_within}
	\end{subfigure}
	~
	\begin{subfigure}[b]{0.23\textwidth}
		\includegraphics[width=\textwidth]{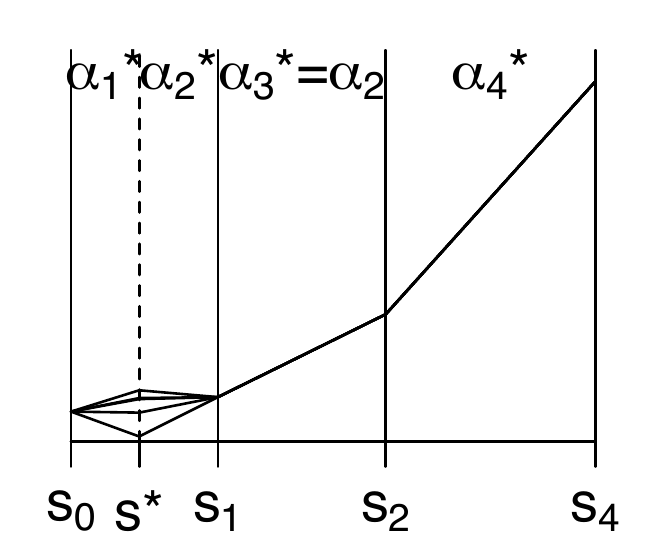}
		\caption{Jump over}
		\label{subfig:MCMC_jump_over}
	\end{subfigure}
	\caption{Proposed state for the separate, jump within and jump over moves are depicted schematically for a hypothetical experiment configuration with $K = 3$.
		In all proposed states, new slopes are proposed to ensure continuity of the ER.
		(a) The current state of the MCMC. $s_3$ is chosen to be updated. (b) A new point $s^*$ is proposed within $(s_2, s_4)$ with the corresponding $\alpha$ parameters constant. (c) Simultaneous move of the experiment configuration and the corresponding $\alpha$'s within $(s_2, s_4)$. (d) The proposed point $s^*$ is located outside the interval $(s_2, s_4)$ and new $\alpha$'s are proposed for the experiment that was split $(s_0, s_1)$, and the experiments that were combined $(s_2, s_4)$.}
	\label{fig:MCMC_moves}
\end{figure}

The experiment configuration and inclusion indicators can be updated separately, or simultaneously. We first describe the separate update of $\mb s$ and $(\tend\alpha^X, \tend\alpha^Y)$, and afterwards we will discuss why occasional simultaneous sampling was deemed necessary.
One of the three moves (separate, jump over, jump within) depicted in \cref{fig:MCMC_moves} is performed at every iteration with probability $0.8, 0.1$, and $0.1$ accordingly.

\paragraph{(separate)}

The experiment configuration and inclusion indicators are updated separately and conditionally on each other. For the update of the experiment configuration $\mb s$, a Metropolis-Hastings step is used \citep{Metropolis1953,Hastings1970} based on the full conditional likelihood \cref{app_eq:factor_conditional}. $k$ is chosen uniformly over $\{1, 2, \dots, K\}$ and $s^* \sim U(s_{k - 1}, s_{k + 1})$ is drawn as shown in \cref{fig:MCMC_moves}(\subref{subfig:MCMC_separate}). Alternatively, $s^*$ could be sampled from a truncated normal distribution centered at $s_k$. If $s^*$ violates $s_{k +1} - s^* \geq d_k$ or $s^* - s_{k - 1} \geq d_{k -1}$, the move is automatically rejected.

\begin{figure}[!t]
	\centering
	\includegraphics[width = 0.4\textwidth]{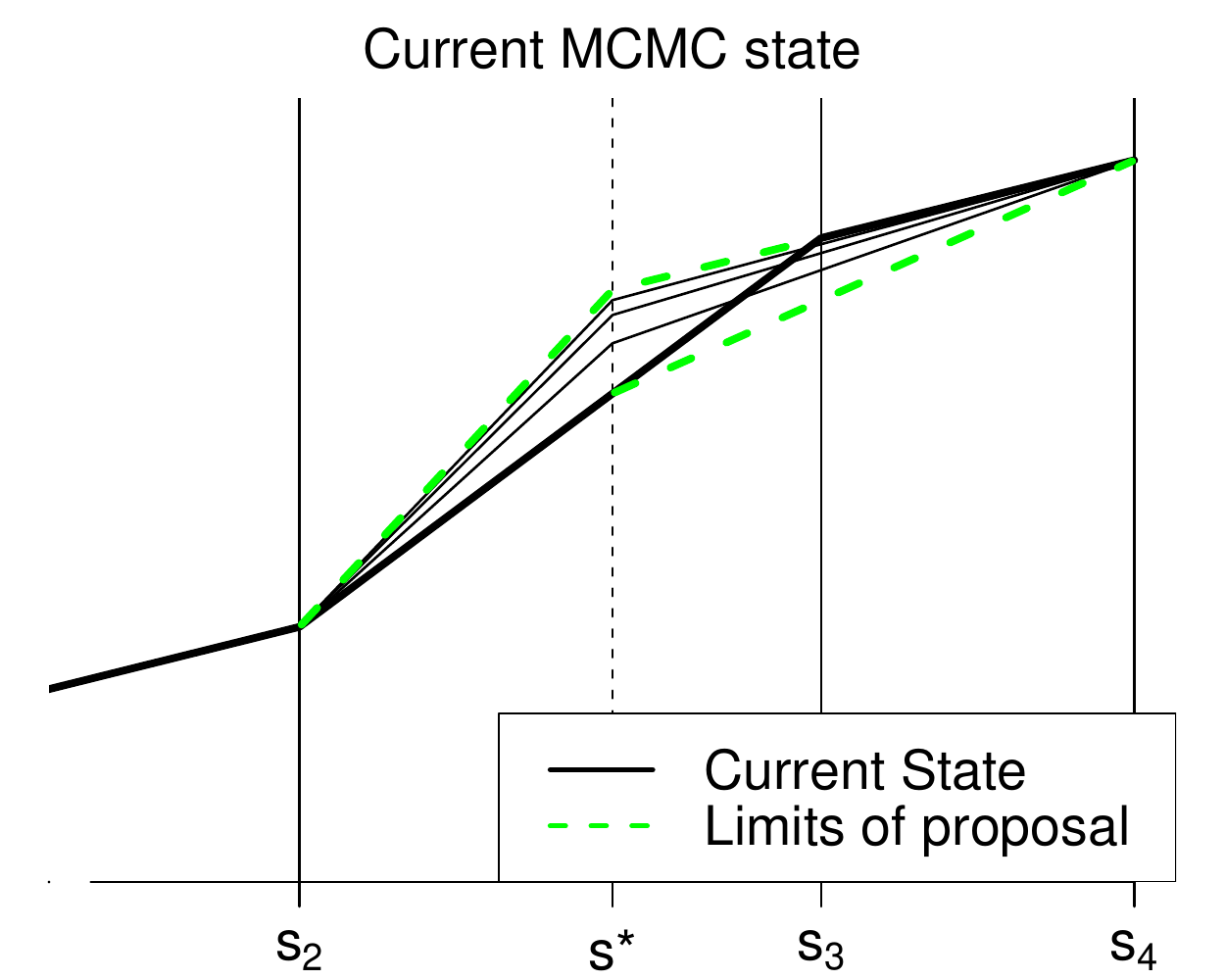}
	\caption{Values of $\beta_k, \beta_{k + 1}$ for the separate move are proposed such that the estimated ER are within the limits shown in dashed green lines. The black solid line correspond to the current state of the ER.}
	\label{app_fig:MCMC_separate_slope}
\end{figure}

Otherwise, the move $\mb s \rightarrow \mb s^* = (s_1, s_2, \dots, s_{k - 1}, s^*, s_{k + 1}, \dots, s_K)$ is proposed with all other parameters (excluding $\tend\beta$) fixed to their current values. Proposing new values of $\tend\beta$ is necessary to ensure that the ER is continuous at the proposed state. All coefficients but $\beta_k, \beta_{k + 1}$ are fixed to their current values, and new values for $\beta_k, \beta_{k + 1}$ are proposed such that the intercepts of the adjacent experiments are also fixed. If $s^* < s_k$, the proposed value $\beta_{k +1}^*$ is sampled from a uniform distribution between the values $\beta_{k+1}$ (current state) and
$$
\tilde \beta_{k+1} = (s_{k + 1} - s^*)^{-1} \big( \delta_{(k + 2)0}^Y - \delta_{k0}^Y - \beta_k(s^* - s_{k-1}) \big),
$$
where $\tilde \beta_{k+1}$ is the slope that would connect the value of the ER at point $s_{k + 1}$ with the value of the ER at point $s^*$ at the current state.
\cref{app_fig:MCMC_separate_slope} shows the the limits of the proposed ER. Based on the sampled value for $\beta_{k+1}^*$, the proposed value for $\beta_k$ is
$$
\beta_k^* = (s^* - s_{k - 1})^{-1} \big(\delta_{(k+2)0}^Y - \delta_{k0}^Y - \beta_{k + 1}^*(s_{k +1} - s^*)\big).
$$
Similarly for $s^* > s_k$ by sampling $\beta_k^*$ from a uniform that has similar properties.

Since the likelihood factorizes as shown in \cref{app_eq:factor_conditional} the likelihood ratio of the Metropolis-Hastings acceptance probability includes terms only for experiments $k, k + 1$. The prior ratio includes terms for the experiment configuration distribution in (\ref{app_eq:prior_experiments}), and the prior for $\beta_k, \beta_{k + 1}$. If a uniform distribution is used to sample $s^*$, the proposal for the cutoffs is symmetric, and the proposal ratio corresponds to the proposal ratio for coefficients $\beta_k, \beta_{k + 1}$. This is equal to $|\beta_{k + 1} - \tilde \beta_{k + 1}| / |\beta_k^* - \tilde \beta_k^*|$, where $\beta_k^*$ is the proposed value and $\tilde\beta_k^*$ is the one boundary of the proposal distribution for $\beta_k$ in the reverse move.

After we accept or reject the move $\mb s\rightarrow \mb s^*$, we update the inclusion indicators based on their full conditional. Let $A^*$ be all parameters but $\alpha_{kj}^X$, $\alpha_{kj}^Y$ and $\delta_{kj}^Y$. For $\alpha \in \{0, 1\}$
\begin{align*}
p(\alpha_{kj}^Y = \alpha |\text{Data}, A^*, \alpha_{kj}^X) = &
\frac{p(\delta_{kj}^Y = 0,\alpha_{kj}^Y = \alpha | \text{Data}, A^*, \alpha_{kj}^X)}
{p(\delta_{kj}^Y = 0 | \alpha_{kj}^Y = \alpha, \text{Data}, A^*, \alpha_{kj}^X)} \\
= & \frac{p(\text{Data}, A^* | \delta_{kj}^Y = 0,\alpha_{kj}^Y = \alpha, \alpha_{kj}^X)
	p(\delta_{kj}^Y = 0,\alpha_{kj}^Y = \alpha | \alpha_{kj}^X)}
{p(\text{Data}, A^* | \alpha_{kj}^X)
	p(\delta_{kj}^Y = 0 | \alpha_{kj}^Y = \alpha, \text{Data}, A^*, \alpha_{kj}^X)} \\
\propto & \frac{p(\delta_{kj}^Y = 0 | \alpha_{kj}^Y = \alpha, \alpha_{kj}^X)
	p(\alpha_{kj}^Y = \alpha |\alpha_{kj}^X)}
{p(\delta_{kj}^Y = 0 | \alpha_{kj}^Y = \alpha, \text{Data}, A^*, \alpha_{kj}^X)} \\
= & \frac{p(\delta_{kj}^Y = 0 | \alpha_{kj}^Y = \alpha)
	p(\alpha_{kj}^Y = \alpha |\alpha_{kj}^X)}
{p(\delta_{kj}^Y = 0 | \alpha_{kj}^Y = \alpha, \text{Data}, A^*)}, \ \alpha \in \{0, 1\},
\numberthis
\label{app_eq:alpha_post}
\end{align*}
where the numerator consists of the product of two prior probabilities, and the denominator consists of the posterior probability that $\delta_{kj}^Y = 0$.
This has been seen previously in a different context \citep{Antonelli2017bayesian}, and consists a computational improvement over previous implementations of this prior distribution that utilized the MC$^3$ algorithm \citep{Madigan1995,Wang2012}.

\paragraph{}
However, sampling the inclusion indicators and experiment configuration separately can lead to slow convergence. For example, consider our simulation scenario where the true experiment configuration is $(2, 4, 7)$, and starting values randomly set to $(0.5, 2, 7)$. Based on the separate move, point $s_1$ is always proposed to be updated between $s_0, s_2 = 2$, which can lead to slow mixing. The jump over and jump within moves are meant to alleviate such issues.

In order to avoid the need of proposing values for the covariates' coefficients and variance terms in the update of the experiment configuration through a simultaneous move, these parameters are integrated out from the data likelihood. Integrating all other parameters out allows us to perform sampling of the experiment configuration without heavy fine tuning of proposal distributions.
In both situations, sampling of $\mb s, \tend{\alpha}^X, \tend{\alpha}^Y, \tend\beta$ is performed using the marginalized likelihood \cref{app_eq:marginal_factorization}:
\begin{align*}
p(\mb s, \tend{\alpha}^X, \tend{\alpha}^Y, \tend\beta | \text{Data}, \delta_{10}^Y)\ \propto \ 
p(\mb Y, \mb X| \mb s, \tend{\alpha}^X, \tend{\alpha}^Y, \tend\beta, \delta_{10}^Y, \mb C)\ 
p(\mb s)\  p(\tend{\alpha}^X, \tend{\alpha}^Y)\ p(\tend\beta).
\end{align*}
Note that all likelihoods in (\ref{app_eq:marginal_factorization}) are marginal densities of linear regression models over the regression coefficients and variance terms with Normal-Inverse Gamma priors. \cite{Raftery1997} provided closed form calculations of this marginal likelihood. However, this calculation requires the inversion of a matrix with dimension equal to the number of observations, and is computationally intensive. Since the marginal likelihood is only used in the calculation of Bayes factors, we approximate the Bayes factors when necessary using the BIC \citep{Raftery1995}.

\paragraph{(jump over)}

This move is designed to alleviate the MCMC issue described above by proposing a simultaneous move of $(\mb s, \tend\alpha^X, \tend\alpha^Y, \tend\beta)$. $k \in \{1, 2, \dots, K\}$ is again chosen uniformly, but now a new location of the experiment configuration $s^*$ is generated uniformly over $(s_0, s_{K + 1}) \setminus [s_{k -1}, s_{k + 1}]$ (necessarily not between $s_k, s_{k + 1}$). The move $\mb s \rightarrow \mb s^* =
(s_1, s_2, \dots, s_{k - 1}, s_{k + 1}, \dots, s_{j - 1}, s^*, s_j, \dots, s_K)$ proposes a combination of experiments $k, k + 1$ and a split in some randomly chosen experiment $j$.
For example, in \cref{fig:MCMC_moves}(\subref{subfig:MCMC_jump_over}), the proposed move splits the first experiment $(s_0, s_1)$ in two $(s_0, s^*), (s^*, s_1)$, and combines the experiments $(s_2, s_3), (s_3, s_4)$.

The inclusion indicators of the unchanged experiments remain to their current values, but new values need to be proposed for the combined or split experiments.
The assignment of proposed values for the inclusion indicators is probabilistic based on their current values, encouraging the inclusion of a covariate in the proposed state to resemble that of the current state. For example, in \cref{fig:MCMC_moves}(\subref{subfig:MCMC_jump_over}) $\alpha_1^*, \alpha_2^*$ should resemble $\alpha_1$, and similarly for $\alpha_4^*$. In this example, covariates are included in experiment 4 with very low, mediocre and very high probability if none, one or both of the original experiments include it. The values chosen for these probabilities were $(0.01, 0.5, 0.99)$ accordingly. Similarly, a variable is proposed to be included in the model of experiments 1 and 2 with low and high probability if the variable was included in the initial model or not. The values chosen were $(0.2, 0.95)$.

Values for $\tend\beta$ are proposed to ensure that the proposed state corresponds to a continuous ER. Unchanged experiments remain the same. Experiments are combined by connecting the edges of the two linear segments, and values of the split experiments are proposed using a normal perturbation of the current value with variance $\sigma^2_{tune}$. \cref{fig:MCMC_moves}(\subref{subfig:MCMC_jump_over}) shows proposed states of the ER.

The move is accepted or rejected with probability equal to the product of the following:
\begin{enumerate}[leftmargin=*]
	\item The likelihood ratio for split and combined experiments approximated using the BIC for the exposure model and the outcome model (regressing $\mb Y^k - (\mathbb{1}, \mb X^k - s_{k - 1}\mathbb{1})(\delta_{k0}^Y, \beta_k)^T$ on $\mb C^k$ without an intercept).
	\item The prior ratio for the experiment configuration (\ref{app_eq:prior_experiments}), the inclusion indicators \cref{eq:BAC_prior}, and the coefficients $\beta_k$ for the combined and split experiments.
	
	\item The proposal ratio for $\mb s$, $\beta_k$ and $(\tend\alpha^X, \tend\alpha^Y)$
	{\small
		\begin{align*}
		\frac{(s_{K + 1} - s_0)-(s_j - s_{j - 1})}
		{(s_{K + 1} - s_0)-(s_{k + 1} - s_{k - 1})}
		\exp\left\{\frac{u^2 - u^{*2}}{2\sigma^2_{tune}} \right\}
		\prod_{\substack{l \in \{0, 1, 2\} \\ m\in \{0, 1\}}}
		(p^c_{lm})^{n^s_{ml} - n^c_{lm}}
		\prod_{\substack{l \in \{0, 1\} \\ m\in \{0, 1, 2\} }}
		(p^s_{lm})^{n^c_{ml} - n^s_{lm}},
		\end{align*}}
	\hspace{-8pt}
	where $p^c_{lm}$ is the probability of proposing $\alpha = m \in\{0, 1\}$ in the combined experiment when $l\in \{0, 1, 2\}$ of the two initial experiments had $\alpha = 1$,
	$p^s_{lm}$ is the probability of proposing $\alpha = 1$ in $m \in \{0, 1, 2\}$ of the two experiments when the initial experiment chosen to be split had $\alpha = l \in \{0, 1\}$, and
	$n^c_{lm}$, $n^s_{lm}$ is the number of times that each event occurred when moving from the current to the proposed state. Lastly, $u$ is the difference of the slope for the experiment that was split from the slope of the first split experiment in the proposed state, and $u^*$ is the difference of the slope in the first of the experiment that is combined from the slope of the combined experiment in the proposed state. 
\end{enumerate}

\paragraph{(jump within)}

This move is similar to the ``jump over'' but maintaining the ordering of the locations in $\mb s$. $k \in \{1, 2, \dots, K\}$ is again chosen uniformly, and a new value $s^*$ is proposed within the interval $(s_{k - 1}, s_{k + 1})$. New values for the coefficients $\beta_k, \beta_{k +1}$ are proposed as in the separate move. New values of the inclusion indicators are also proposed for the experiments $k, k + 1$. In fact, $C_j$ is proposed to be included in the outcome model of an experiment with high probability if both current models include it, mediocre probability if only one of the models include it, and low probability if none of the models include it. Similarly for the inclusion indicators of the exposure model. The acceptance probability of this move is similar to the one described above, and is omitted here. 
\cref{fig:MCMC_moves}(\subref{subfig:MCMC_jump_within}) depicts random draws for proposed ER states.

\subsection{MCMC convergence}
\label{app_subsec:MCMC_convergence}

Due to the update of the experiment configuration, commonly used convergence diagnostics such as trace plots are not appropriate since parameters (e.g., $\beta_k$) may correspond to a different range of exposure values at different iterations. Therefore, convergence must be examined in the context of quantities that are detached from the experiment configuration.

One quantity that we use for convergence inspection is the mean exposure response curve calculated over a set of exposure values within the exposure range. Such a set might be an equally spaced grid of points over the interval $(s_0, s_{K + 1})$, denoted by $\mathcal{G}$.
For each value $x \in \mathcal{G}$ and MCMC iteration $t$, identify the experiment $k = k_t(x)$ that $x$ belongs to.
Then, for observation $i$  calculate the expected response at value $x$, by defining $\tilde{w}_i(x) = (1, x, C_{i1}, \dots, C_{ip})^T$ and calculating $\widehat{Y}_{it}(x) = \tilde{w}_i(x) ^T \gamma_{kt}$ where $\gamma_{kt}$ is the posterior sample of $(\delta_{k0}^Y, \beta_k, \delta_{k1}^Y, \dots, \delta_{kp}^Y)^T$ in iteration $t$.
Finally, the $t$-posterior sample of the mean response at point $x\in \mathcal{G}$ is the average of the expected responses over the individuals in the sample $\widehat{Y}_t(x) = \frac1n \sum_{i = 1}^n \widehat Y_{it}(x)$.

Convergence could be examined by visual inspection of trace plots of $\widehat Y(x)$ for all $x \in \mathcal{G}$. Based on multiple chains of the MCMC, we calculate the potential scale reduction factor (PSR) for the mean response at every point $x\in \mathcal{G}$ \citep{Gelman1992}. We consider that the MCMC has converged if  $|\text{PSR} - 1| < c$ for all $x \in \mathcal{G}$.
An alternative quantity based on which MCMC convergence can be examined is $\hat\Delta(x) = \beta_{k_t(x)}$.

\section{Simulating data with differential confounding at different exposure levels}
\label{app_sec:sim_mech}

In simulation studies, data are most often simulated in the following order: covariates $C_1, C_2, \dots, C_p$, exposure $X$ given a subset of $C_1, C_2, \dots, C_p$, and outcome $Y$ given $X$ and a potentially different subset of $C_1, C_2, \dots, C_p$.
Data with differential confounding at different exposure levels could imply, in its most generality, that the exposure $X$ is generated with different predicting variables at different exposure levels. Generating data with such structure is complicated since the actual $X$ values define the exposure level that an observation belongs to, and the exposure level in which an observation belongs to defines the set of predictors.
For that reason, instead of following the $\mb C, X | \mb C$ approach to data simulation, we generate the exposure values $X$ first, and $\mb C$ is generated conditional on $X$, ensuring that the target experiment-specific mean and variance of $X, \mb C$, and correlation of all variables remain the same, as if the data were generated with the typical $\mb C, X | \mb C$ order.
Generating the outcome with different predictors at different exposure levels is straightforward by including terms of the form $\delta^*_j C_j I(X \geq s_k)$, or by using a separate outcome model within each experiment. In all situations, one should ensure that data are generated in such a way that the true ER is continuous.

\subsection{The ``target'' data generating mechanism}
\label{app_subsec:target_generate}

Given $K, \mb s$, we would like the exposure $X$ to be generated such as $E(X)$ and $Var(X)$ are controllable quantities, since they are closely related to the exposure range of each experiment, and we would like to ensure that simulation results are not driven by the inherit variability in $X$. Furthermore, we would like to ensure that $Var(C_j)$ is approximately the same across experiments and across covariates, such that the the magnitude of $\delta_{kj}^Y$ has similar interpretation in terms of correlation.

As discussed above, data $(X, \mb C)$ are usually generated in the order $\mb C$ followed by $X | \mb C$, using a model for which $E(X|\mb C)  = \delta_0 + \sum_{j = 1}^p \delta_j C_p$. Instead of setting target values for $\delta_j$, we set target correlations $Cor(X, C_j)$ and calculate the $\delta_j$'s that correspond to these correlations. (The reverse is also possible but requires ensuring that that $Var(X) \geq \sum_{j = 1}^p \delta_j^2 Var(C_j)$.)
We require that $E(C_j | X = x)$ is continuous in $x$ to ensure that the joint distribution $(X, C_j)$ is realistic, and does not have ``jumps'' at the points of the experiment configuration.

Based on the above, the following represent target (controllable) quantities of our data generation:
\begin{itemize}
	\item $Var(X), E(X)$ are fixed,
	\item Within each experiment $C_j$ are independent random variables with known variance,
	\item $Cor(C_j, X)$ are fixed and $\delta_j$ can be calculated, using
	$\displaystyle Cor(X, C_j) = \delta_j \sqrt{\frac{Var(C_j)}{Var(X)}}$,
	\item The function $E(C_j | X = x)$ is continuous in $x$.
\end{itemize}

Ensuring that $E(C_j | X= x)$ is continuous in $x$ across experiments is performed in the following way: Given $Var(X)$, a model for $\mb C | X$ that gives rise to data with the target $Var(C_j), Cor(X, C_j)$ is considered. The variance-covariance targets do not impose any restrictions on the model intercept. For the first experiment, the intercept can be chosen arbitrarily, and for the subsequent experiments intercepts are chosen to ensure that $\lim_{t \rightarrow x^-}E(C_j | X = t) = \lim_{t \rightarrow x^+}E(C_j | X = t)$ at all points $x$.

\subsection{Generating the data set maintaining target quantities}
\label{app_subsec:generate}

As discussed in \S\ref{app_subsec:target_generate}, $Cor(X, C_j)$, $Var(C_j)$, and $Var(X)$ are considered known, from which we can derive $Cov(X, C_j)$.
We generate data with the following order:
\begin{enumerate}
	\item $X$ is generated from a distribution with mean $E(X)$, and variance $Var(X)$. In our simulations $X$ is uniform over the exposure range.
	\item Taking advantage of the laws of the multivariate normal distribution we generate
	\begin{align*}
	\mb C | X \sim & MVN_p(\bar\mu, \bar\Sigma), \text{ where}\\
	\bar \mu = & E(\mb C) + \frac{Cov(\mb C, X)}{Var(X)} (X - EX),
	\text{ and } \\
	\bar\Sigma = & V(\mb C) -
	\frac{1}{Var(X)} Cov(\mb C, X) Cov(\mb C, X)^T,
	\end{align*}
	where $Cov(\mb C, X) = (Cov(C_1, X), Cov(C_2, X), \dots, Cov(C_p, X))^T$, and $V(\mb C)$ is a diagonal $p\times p$ matrix with entries $Var(C_j), j = 1, 2, \dots, p$.
	\item The marginal means of each variable $C_j$ within each experiment is calculated by ensuring that the function $E(C_j | X = x)$ which corresponds to the $j^{th}$ entry of the vector $\bar\mu$ is continuous at the points of experiment change.
	\item Covariates $C_j$ are subtracted their overall mean.
\end{enumerate} 

A simple linear regression form is used to generate the outcome within each experiment. In experiment $k$, the outcome is generated from 
$Y | X, \mb C \sim N(\xi_{k0} + \xi_{k1}\phi(X) + \sum_{j = 1}^p \xi_{k(j+1)} C_j, \sigma^2_{k, Y})$, where $\phi()$ is a continuous function, and the residual variance $\sigma^2_{k,Y}$ is set equal across $k$.
We ensure that the true ER function $E(Y | X)$ is continuous in $X$ by appropriately setting the intercept values $\xi_{k0}$. The intercept in experiment 1 is decided, and for each experiment onwards we set $\xi_{k0}$ such that
\begin{align*}
& \lim_{x \rightarrow s_k^-} E[Y | X = x] = \lim_{x \rightarrow s_k^+} E[Y | X = x ]
\iff \xi_{(k + 1)0} = \xi_{k0} + (\xi_{k1} - \xi_{(k+1)1})\phi(s_k).
\end{align*}

\end{document}